\documentclass[fleqn,usenatbib,useAMS]{mnras}

\usepackage{graphicx}	
\usepackage{amsmath}	
\usepackage{amssymb}	
\usepackage{multicol}        
\usepackage{adjustbox}
\usepackage{bm}		
\usepackage{tikz,xcolor,hyperref}
\usepackage{soul}

\definecolor{lime}{HTML}{A6CE39}
\DeclareRobustCommand{\orcidicon}{
	\begin{tikzpicture}
	\draw[lime, fill=lime] (0,0) 
	circle [radius=0.16] 
	node[white] {{\fontfamily{qag}\selectfont \tiny ID}};
	\draw[white, fill=white] (-0.0625,0.095) 
	circle [radius=0.007];
	\end{tikzpicture}
	\hspace{-2mm}
}

\foreach \x in {A, ..., Z}{\expandafter\xdef\csname orcid\x\endcsname{\noexpand\href{https://orcid.org/\csname orcidauthor\x\endcsname}
			{\noexpand\orcidicon}}
}

\usepackage[T1]{fontenc}
\usepackage{ae,aecompl}

\usepackage{newtxtext,newtxmath}

\title[High energy gamma-ray sources in the VVV survey]{High energy gamma-ray sources in the VVV survey - II. The AGN counterparts}

\author[Donoso et al.]
{
\parbox[t]{\textwidth}{Laura G. Donoso\orcidK$^{1}$, Ana Pichel\orcidE$^{2}$, Laura D. Baravalle\orcidA$^{3,4}$, M. Victoria Alonso\orcidD$^{3,4}$, Eduardo O. Schmidt\orcidB$^{3,4}$, Dante Minniti\orcidF$^{5,6,7}$, Nicola Masetti\orcidG$^{8,5}$, 
Leigh C. Smith\orcidI$^{9}$, Philip W. Lucas\orcidJ$^{10}$, Carolina Villalon\orcidH$^{3}$, Adri\'an C. Rovero$^{2}$ \& Georgina Coldwell$^{11}$}
\vspace*{6pt} \
\\
$^{1}$ Instituto de Ciencias Astron\'omicas, de la Tierra y del Espacio (ICATE, CONICET), C.C. 467, 5400, San Juan, Argentina.\\
$^{2}$ Instituto de Astronom\'ia y F\'isica del Espacio (IAFE, CONICET-UBA), Ciudad Aut\'onoma de Buenos Aires, Argentina.\\
$^{3}$ Instituto de Astronom\'ia Te\'orica y Experimental (IATE, CONICET-UNC), Laprida 854, C\'ordoba, Argentina.\\
$^{4}$ Observatorio Astron\'omico de C\'ordoba, Universidad Nacional de C\'ordoba, C\'ordoba, Argentina.\\
$^{5}$ Instituto de Astrof\'isica, Facultad de Ciencias Exactas, Universidad Andr\'es Bello, Av. Fernandez Concha 700, Las Condes, Santiago, Chile.\\
$^{6}$ Vatican Observatory, V00120 Vatican City State, Italy.\\
$^{7}$ Departamento de F\'isica, Universidade Federal de Santa Catarina, Trinidade 88040-900, Florianopolis, Brazil.\\
$^{8}$ INAF - Osservatorio di Astrofisica e Scienza dello Spazio, via Piero Gobetti 101, I-40129 Bologna, Italy.\\
$^{9}$ Institute of Astronomy, University of Cambridge, Madingley Road, Cambridge CB3 0HA, UK.\\
$^{10}$ Centre for Astrophysics, University of Hertfordshire, College Lane, Hatfield AL10 9AB, UK.\\
$^{11}$ Departamento de Geof\'isica y Astronom\'ia, Facultad de Ciencias Exactas, F\'isicas y Naturales (CONICET-UNSJ), San Juan, Argentina.\\
}

\date{\today}

\begin{document}
\label{firstpage}
\pagerange{\pageref{firstpage}--\pageref{lastpage}}
\maketitle

\begin{abstract}
We identified Active Galactic Nuclei (AGN) candidates as counterparts to unidentified gamma-ray sources (UGS) from the Fermi-LAT Fourth Source Catalogue at lower Galactic latitudes. Our methodology is based on the use of near- and mid-infrared photometric data from the VISTA Variables in the V\'ia L\'actea (VVV) and Wide-field Infrared Survey Explorer (WISE) surveys. The AGN candidates associated with the UGS occupy very different regions from the stars and extragalactic sources in the colour space defined by the VVV and WISE infrared colours. We found 27 near-infrared AGN candidates possibly associated with 14 Fermi-LAT sources using the VVV survey. We also found 2 blazar candidates in the regions of 2 Fermi-LAT sources using WISE data.  There is no match between VVV and WISE candidates. We have also examined the K$_\mathrm{s}$ light curves of the VVV candidates and applied the fractional variability amplitude ($\mathrm{\sigma_{rms}}$) and the slope of variation in the  K$_\mathrm{s}$ passband to characterise the near-infrared variability. This analysis shows that more than 85\% of the candidates have slopes in the K$_\mathrm{s}$ passband $ > 10^{-4}$ mag/day and present $\mathrm{\sigma_{rms}}$ values consistent with a moderate variability. This is in good agreement with typical results seen from type-1 AGN. The combination of YJHK$_\mathrm{s}$ colours and K$_\mathrm{s}$ variability criteria was  useful for AGN selection, including its use in identifying counterparts to Fermi $\gamma$-ray sources.
\end{abstract}




\section{Introduction}

Since its launch in June 2008, the Fermi Large Area Telescope (\citealt{Atwood2009}, Fermi-LAT) has revolutionised our view of the $\gamma$-ray sky above 100 MeV. 
The Fermi-LAT offers a significant increase in sensitivity, improved angular resolution and nearly uniform sky coverage, making
it a powerful tool for the detection and characterisation of large numbers of $\gamma$-ray sources. 
The Fermi Fourth Source Catalogue (\citealt{Abdollahi2020}, 4FGL), based on the first 8 years of data from the mission, 
lists 5064 sources in the energy range 50 MeV to 1 TeV. 
Out of these sources, 1336 (26.4\%) sources do not have even a reliable association with 
sources detected at other wavelengths; we will henceforth label them as Unassociated Gamma-ray Sources (UGS). More
than 3130 of the identified or associated sources are active galaxies of the blazar class,
and 239 are pulsars. 

The positions of $\gamma$-ray sources listed in the Fermi-LAT catalogues are reported with
their associated uncertainty represented by an elliptical region.
The Fermi-LAT $\gamma$-ray catalogues provide the semi-major and semi-minor axes of the ellipses together with the
positional angle at 68\% and 95\% level of confidence. The principal reason for the difficulty of finding counterparts to high-energy $\gamma$-ray sources has been the large positional errors in their measured locations, a result of
the limited photon statistics and angular resolution of the $\gamma$-ray observations and the bright diffuse
$\gamma$-ray emission from the Milky Way (MW). Therefore, the UGS represent one of the biggest challenges in $\gamma$-ray
astrophysics \citep[e.g.,][]{Thompson2008}. 
The key to finding plausible counterparts to the unidentified Fermi-LAT sources is the cross-check with observations  
at one or more wavelengths, such as radio observations \citep[e.g.,][]{Hovatta2014, Shinzel2015}, infrared observations \citep[e.g.,][]{Raiteri2014} and in the
sub-millimeter range \citep[e.g.,][]{LeonTavares2012, LopezCaniego2013}. 
Additional X-ray studies have also been 
carried out with Chandra and Suzaku have been useful
in particular when performed in the crowded region of the Galactic plane \citep[e.g.,][]{Maeda2011, Cheung2012}. Optical spectroscopic identification of Fermi sources has been
addressed previously to search for counterparts \citep[e.g.,][]{Paggi2014,PenaHerazo2021,GarciaPerez2023}.  
In addition, the properties of the $\gamma$-ray sources can be used as a statistical set to perform a multivariate analysis.
This is a classification strategy to find  plausible counterparts at other wavelengths for sources that remain unassociated \citep[e.g.,][]{Hassan2013, Doert2014}.  

Active Galactic Nuclei (AGN) represent an astronomical phenomenon that emit extremely high-energy radiation, as demostrated by \citet[][]{Urry1995} and \citet[][]{Padovani2017}.  Since their discovery many decades ago, research has been conducted at various frequencies unveiling the diverse manifestations of AGN phenomena, observed from radio to $\gamma$-rays. This has resulted in an extensive and captivating assortment of classifications. Among the distinct classes of AGN are type-1 and type-2 AGN, blazars subdivided in BL Lacertae  and Flat Spectrum Radio Quasars (FSQR), alongside other classifications (see, \citealt{Stickel1991, Stocke1991}). 
The AGN unification scheme, as proposed by \citet{Antonucci1993}, offers a comprehensive representation of AGN phenomena, including elements such as black holes, discs, torus, clouds, and jets. This model explains how orientation effects, different accretion powers, and black hole spin parameters can account for the wide array of AGN types. Furthermore, AGN typically exhibit variations in their emissions \citep{Edelson2002, Sandrinelli2014, Husemann2022}. The extent of this variability differs according to the type of AGN and is generally more pronounced, with higher amplitudes in blazars compared to type-1 AGN \citep[e.g.,][]{Ulrich1997, Mao2021, Baravalle2023}.  

In recent years, the population of known AGN has substantially grown thanks to new surveys and catalogues \citep[e.g.,][]{Veron2010, Rembold2017, Donacimento2019}. Nevertheless, the number of AGN observed at lower Galactic latitudes, obscured by dense regions belonging to our Galaxy,
remains limited \citep[e.g.,][]{Edelson2012, Pichel2020}. Recently, \citet{Fu2021, Fu2022} explored the Galactic bulge regions in search for quasars (QSO) at lower latitudes. Employing machine learning techniques, they identified and confirmed 204 QSO candidates at ($\mathrm{\mid b\mid < 20^{\circ}}$) based on spectroscopic measurements. \citet{Ackermann2012} reported a significant excess of unassociated sources at $\mathrm{\mid b\mid < 10^{\circ}}$, where catalogues of AGN are incomplete. 
Hence the fraction of sources associated with AGN decreases in this sky area. 

Extragalactic objects located behind the Milky Way are difficult to identify and detect due to the significant amount of gas, dust, and stars present at low Galactic latitudes \citep[e.g.,][]{Kraan2000,Baravalle2018,Baravalle2021}. 
In this context, observations carried out at near-infrared wavelengths minimise the effects of interstellar extinction in these regions in comparison with optical passbands. Although the density of foreground sources is greater in the near-infrared, the reduced foreground extinction can reveal different physical processes. Studying these unknown MW regions at low Galactic latitudes, which  are usually obscured at visible wavelengths, presents a challenging task. 
The first near-infrared survey in these regions was the Two Micron All Sky Survey \citep[][2MASS]{Skrutskie2006}. Later, the ESO Public Surveys, the VISTA Variables in the V\'ia L\'actea \citep[][VVV]{Minniti2010} and its extension, the VVVX have been mapping the K$_\mathrm{s}$-passband variability of stars in the entire MW bulge and disc. The main scientific goal was to gain more insight into the inner MW's origin, structure, and evolution.
The VVV survey included the acquisition of ZYJHK$_\mathrm{s}$ images whereas VVVX was restricted to the JHK$_\mathrm{s}$ passbands, increasing significantly the coverage area (see Table 1 in \citealt[]{Daza2023}).
Thousands of new galaxies and galaxy associations have been discovered using the photometric data from VVV and VVVX surveys \citep[e.g.,][]{Amores2012, Baravalle2019, Coldwell2014, Galdeano2021, Soto2022, Daza2023}. The VVV near-infrared galaxy catalogue (\citealt{Baravalle2021}, VVV NIRGC) is the final catalogue of part of the Southern Galactic disc using the colour criteria and the visual inspection to identify 5554 galaxies.  Only 45 of these galaxies were previously known.  \cite{Pichel2020} studied for the first time the active galaxies in these regions using a combination of near-infrared (NIR) and mid-infrared (MIR) data. The Wide-field Infrared Survey Explorer \citep[][WISE]{Wright2010} is an ideal mission for identifying a very large number of AGN across the full sky. Additionally, \cite{Baravalle2023} reported four AGN candidates at very low Galactic latitudes ($\mathrm{\mid b\mid < 2^{\circ}}$) using this combination of VVV and WISE surveys. Also, these sources presented variability in the K$_\mathrm{s}$ light curves reported in the VIVACE catalogue \citep{Molnar2021}. 

The infrared (IR) emission of AGN can be of thermal and non-thermal origin. 
In the case of radio-loud AGN,  specifically blazar subtypes, the non-thermal character of the IR radiation is produced by the synchrotron emission of relativistic electrons within the jet. Radio continuum emission is also associated with these jets.  On the other hand, in radio-quiet objects such as Seyfert galaxies, most of the radiated energy is dominated by thermal emission from the accretion disc, which is formed around the central black hole \citep[e.g.,][]{Shakura1973}. The light of the accretion disc is absorbed by the ``dust torus'' (see, \citealt{Netzer2015}) and re-emitted in the infrared. The emission of torus and accretion disc dominate the AGN spectral energy distribution (SED) at wavelengths longer than $\sim$1 $\mu$m up to a few tens of microns, giving the AGN distinctive red mid-IR colours \citep[e.g.][]{Stern2005, Richards2006, Assef2010}. Therefore, IR passbands are well suited to identify AGN, as their SEDs are very different from those of stars and inactive galaxies. \citet{Chen2005} studied the colour distribution of a sample of blazars and normal galaxies using the 2MASS archival data. The main results from these observations are as follows: (1) the distribution of colours of blazars, in the J-H-K$_\mathrm{s}$ colour-colour
diagram, occupy a region centered at the position (0.7; 0.7), and (2) about 30\% of the blazars show NIR colours indicating a possible influence from the host galaxy. Such contamination is not
present at MIR wavelengths. Using WISE magnitudes, \cite{DAbrusco2012} discovered that blazars emitting in $\gamma$-rays were clearly distinguished from other
classes of galaxies and/or AGN and/or Galactic sources. 
Fermi-LAT blazars inhabit different
regions in the colour-colour diagrams (CCD)
because they are dominated by non-thermal
emission in the mid-IR. This two-dimensional region in the MIR CCD [3.4]-[4.6]-[12]-[22] $\mu$m was originally indicated as the
WISE Gamma-ray Strip \citep[][WGS]{DAbrusco2012}, and the
method was improved in the WISE locus of gamma-ray blazars in  \citet{Dabrusco2013, Dabrusco2014}. 
\citet{Massaro2016} also showed
that the Fermi-LAT blazars are located in specific regions both in NIR and MIR CCD, clearly separated from other extragalactic sources.
\citet{Stern2012} investigated the power of WISE to identify AGN based solely on the [3.4] and [4.6] magnitudes. The selection criteria of [3.4]-[4.6] $>$ 0.8 mag and [4.6] $<$ 15.05 mag produced an AGN sample with a contamination of only 5\%. Following this, \cite{Assef2018} presented two additional colour criteria in their AGN sample: [3.4]-[4.6] $>$ 0.5 mag and [3.4]-[4.6] $>$ 0.77 mag, with a 90\% and 75\% completeness.

The main goal in this study is to identify, at lower Galactic latitudes, unidentified 4FGL sources with NIR and MIR counterparts using the VVV and WISE surveys,
respectively.  The paper is organised as follows. Section~\ref{Data} presents the data which includes the different samples of high energy sources together with the NIR and MIR photometry used in this study.  
The applied methodology to detect the counterparts is also discussed including colour-magnitude and colour-colour diagrams using VVV and WISE surveys, and the VVV K$_\mathrm{s}$ light curves of the near-IR sources and the variability
analysis. Section~\ref{sec:results}
shows the diagrams for the Fermi-LAT source regions with VVV candidates and the analysis of the light curves using the near-IR data. Diagrams with the WISE candidates using mid-IR data are also shown. Section~\ref{sec:final} presents a summary of the main results.

\section{Data and methodology}
\label{Data}

\subsection{The samples of high energy gamma-ray sources}
\label{Sample}

\begin{figure*}
\includegraphics[width=100mm,height=80mm]{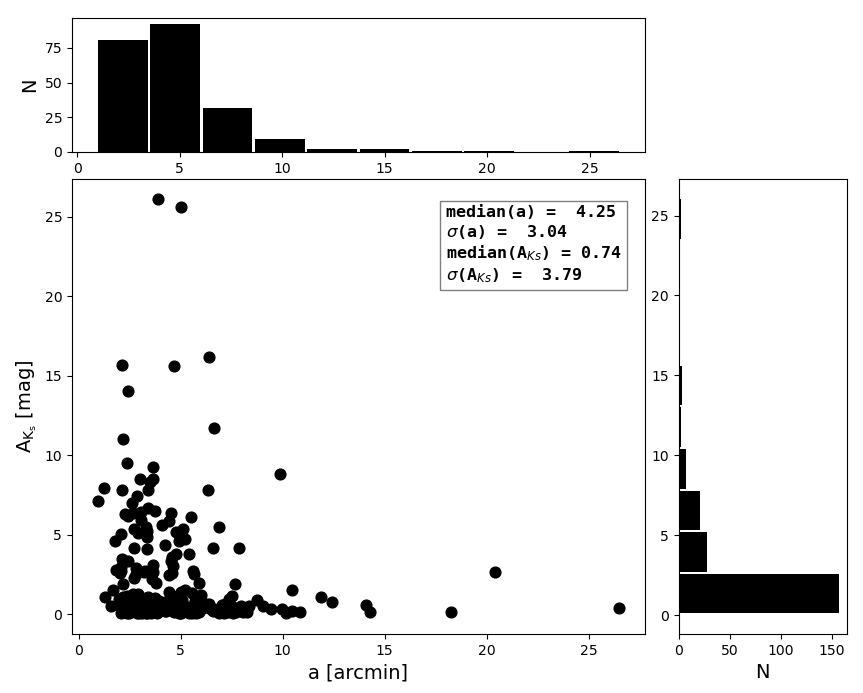}
\caption{Plot resuming the interstellar extinctions in the K$_\mathrm{s}$ passband and uncertainties in the positions of the 4FGL sources in the VVV region.}
\label{f00}
\end{figure*}

At lower Galactic latitudes, we have found 221 4FGL sources in the bulge and disc regions covered by the VVV survey without any previous source associations at any wavelengths.
Figure~\ref{f00} shows the distributions of interstellar K$_\mathrm{s}$ extinctions (A$_\mathrm{Ks}$) in magnitudes and uncertainties in the positions of the Fermi-LAT sources as the semi-major axis ($\mathrm{a}$) in arcmin of the error ellipse at 95\% confidence level for the  221 UGS.  The median values are A$_\mathrm{Ks}$ = 0.74 $\pm$ 3.79 mag and $\mathrm{a}$ = 4.25 $\pm$ 3.04 arcmin.  

According to the distributions of the interstellar extinctions and semi-major axis of the Fermi-LAT uncertainties, we choose to analyse sources in regions with lower interstellar extinctions (A$_\mathrm{Ks} < $ 1.2 mag). Taking this into account, our sample comprises 78 UGS. We defined
three subsamples: the A subsample which contains 13 UGS with $\mathrm{a} < 2.5 $ arcmin; the B subsample that contains 12 sources with 2.5  $\leq \mathrm{a} < 3.0 $ arcmin and the C subsample that contains 53 sources with 3.0  $\leq \mathrm{a} < 5.0 $ arcmin. Tables~\ref{tab1}, ~\ref{tab2} and ~\ref{tab3} show the positions of the UGS for the three subsamples, respectively.  

\begin{figure*}
\includegraphics[width=0.9\textwidth]{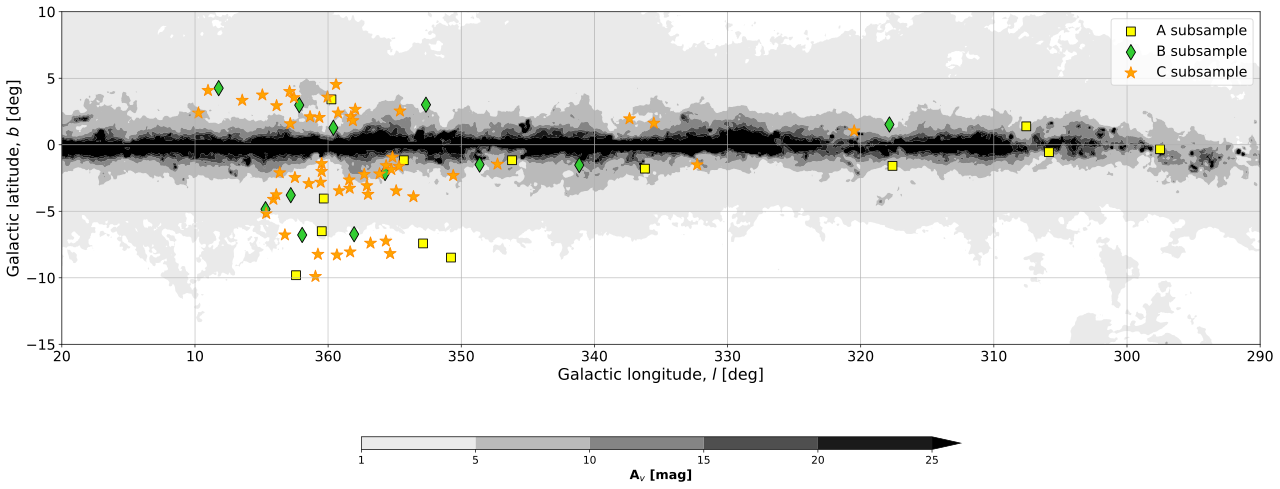}
\caption{The distribution of the 78 UGS in the VVV region using different symbols for A, B and C subsamples. The A$_\mathrm{V}$ iso-contours at 5, 10, 15, 20, 25 mag derived from the extinction maps of \citet{Schlafly2011} are superposed.}
\label{f01}
\end{figure*}

Figure~\ref{f01} shows the distribution in Galactic coordinates of the 78 UGS over the region covered by the VVV survey. The samples studied are highlighted as yellow squares (A subsample), green diamonds (B subsample) and orange stars (C subsample). Also the  coloured UGS are over plotted on the spatial distribution of A$_V$ interstellar extinction derived from the extinction map of \citet{Schlafly2011}. The contours of the different levels correspond to 5, 10, 15, 20, 25 mag. There are 14 UGS located in the Southern disc and 64 in the bulge.  The disc (bulge) UGS are  6, 3 and 5 (7, 9 and 48) in the A, B and C subsamples, respectively.  

\begin{table*}
\caption{Fermi-LAT sources of our sample with low positional uncertainties (the A subsample). Column (1) lists the internal identification used in this work; columns (2) to (5), the 4FGL identification, the J2000 coordinates and the semi-major axis of Fermi-LAT error ellipse, $\mathrm{a}$, at 95\% confidence level in arcmin taken from 4FGL, and columns (6) and (7), the VVV tile identification and the interstellar extinction in the K$_\mathrm{s}$ passband at the source position, respectively.}
 \label{tab1}
 \begin{tabular}{llcccccl}
  \hline
 ID & 4FGL ID & R.A. & Decl. & $\mathrm{a}$ & Tile ID & A$_\mathrm{Ks}$\\
  & &  (J2000)  & (J2000) &  [arcmin]     &         & [mag]   \\    
\hline
A1 & 4FGLJ1203.9-6242 & 12:03:56.23 & -62:42:34.2 & 1.272 & d040 & 1.0926\\
A2 & 4FGLJ1317.5-6316 & 13:17:31.30 & -63:16:43.0 & 2.154 & d046 & 1.0737\\
A3 & 4FGLJ1329.9-6108 & 13:29:56.98 & -61:08:26.2 & 1.602 & d123 & 0.5410\\
A4 & 4FGLJ1456.7-6050 & 14:56:45.12 & -60:50:19.3 & 1.986 & d016 & 0.5663\\
A5 & 4FGLJ1640.3-4917 & 16:40:18.63 & -49:17:19.3 & 2.304 & d029 & 0.9213\\
A6 & 4FGLJ1712.9-4105 & 17:12:56.78 & -41:05:43.4 & 2.352 & d036 & 1.1137\\
A7 & 4FGLJ1731.9-2719 & 17:31:55.70 & -27:19:45.8 & 2.358 & b375 & 0.5330\\
A8 & 4FGLJ1736.1-3422 & 17:36:06.22 & -34:22:37.9 & 2.004 & b315 & 0.9297\\
A9 & 4FGLJ1758.7-4109 & 17:58:47.20 & -41:09:10.1 & 2.082 & b215 & 0.0611\\
A10 & 4FGLJ1759.1-3849 & 17:59:10.34 & -38:49:18.1 & 2.478 & b230 & 0.0767\\
A11 & 4FGLJ1802.4-3041 & 18:02:27.52 & -30:41:57.1 & 2.160 & b277 & 0.2656\\
A12 & 4FGLJ1812.8-3144 & 18:12:52.68 & -31:44:37.7 & 2.250 & b250 & 0.1417\\
A13 & 4FGLJ1830.8-3132 & 18:30:48.79 & -31:32:11.0 & 2.394 & b209 & 0.0628\\
\hline
\end{tabular}
\end{table*}

\begin{table*}
\caption{Fermi-LAT sources  of our sample  with intermediate positional uncertainties (the B subsample). The column description is the same of Table~\ref{tab1}. }
 \label{tab2}
 \begin{tabular}{llcccccl}
  \hline
 ID & 4FGL ID & R.A. & Decl. & $\mathrm{a}$ & Tile ID & A$_\mathrm{Ks}$\\
  & &  (J2000)  & (J2000) &  [arcmin]     &         & [mag]   \\    
\hline
B1 & 4FGLJ1447.4-5757 & 14:47:25.22 & -57:57:04.0 &   2.526 & d130 & 0.6406 \\
B2 & 4FGLJ1657.7-4520 & 16:57:47.52 & -45:20:17.5 &   2.748 & d032 & 1.1385 \\ 
B3 & 4FGLJ1714.9-3324 & 17:14:57.14 & -33:24:48.6 &   2.946 & b370 & 0.4727 \\ 
B4 & 4FGLJ1721.7-3917 & 17:21:43.61 & -39:17:51.0 &   2.886 & d037 & 0.7130 \\ 
B5 & 4FGLJ1739.3-2531 & 17:39:20.79 & -25:31:34.3 &   2.904 & b377 & 0.6179 \\ 
B6 & 4FGLJ1739.7-2836 & 17:39:44.38 & -28:36:08.3 &   2.826 & b347 & 0.8319 \\
B7 & 4FGLJ1743.6-3341 & 17:43:38.55 & -33:41:38.0 &   2.694 & b302 & 0.6330 \\ 
B8 & 4FGLJ1748.2-1942 & 17:48:17.11 & -19:42:23.4 &   2.952 & b395 & 0.3098 \\  
B9 & 4FGLJ1806.9-2824 & 18:06:58.78 & -28:24:56.9 &   2.832 & b279 & 0.2373 \\  
B10 & 4FGLJ1808.4-3358 & 18:08:24.43 & -33:58:54.1 &   2.946 & b248 & 0.1017 \\
B11 & 4FGLJ1815.2-2715 & 18:15:15.15 & -27:15:26.6 &   2.952 & b267 & 0.1562 \\ 
B12 & 4FGLJ1817.2-3035 & 18:17:13.82 & -30:35:03.1 &   2.850 & b251 & 0.0794 \\  
  \hline
\end{tabular}
\end{table*}

\begin{table*}
\caption{Fermi-LAT sources of our sample with large positional uncertainties (the C subsample). The column description is the same as in  Table~\ref{tab1}.}
 \label{tab3}
 \begin{tabular}{llcccccl}
  \hline
ID & 4FGL ID & R.A. & Decl. & $\mathrm{a}$ & Tile ID & A$_\mathrm{Ks}$ \\
 &  &  (J2000)  & (J2000) &  [arcmin]     &         & [mag]   \\    
\hline
C1 & 4FGLJ1506.5-5708 & 15:06:32.64 & -57:08:50.6 &  3.306 & d094 & 1.0008 \\  
C2 & 4FGLJ1622.0-5157 & 16:22:02.38 & -51:57:05.0 &  4.500 & d026 & 0.6612 \\ 
C3 & 4FGLJ1622.5-4726 & 16:22:32.38 & -47:26:05.6 &  4.992 & d142 & 0.9888 \\ 
C4 & 4FGLJ1628.6-4553 & 16:28:36.58 & -45:53:47.4 &  4.290 & d144 & 0.5998 \\ 
C5 & 4FGLJ1717.5-4022 & 17:17:34.63 & -40:22:43.7 &  3.720 & d037 & 0.7776 \\
C6 & 4FGLJ1722.1-3205 & 17:22:06.67 & -32:05:16.4 &  3.330 & b358 & 0.6575 \\
C7 & 4FGLJ1726.8-2659 & 17:26:50.79 & -26:59:56.4 &  4.140 & b389 & 0.4855 \\ 
C8 & 4FGLJ1730.3-2913 & 17:30:18.44 & -29:13:48.7 &  4.692 & b360 & 0.5339 \\
C9 & 4FGLJ1730.8-3806 & 17:30:53.28 & -38:06:47.2 &  3.960 & b299 & 0.8399  \\
C10 & 4FGLJ1732.0-2659 & 17:32:00.02 & -26:59:20.4 &  4.398 & b375 & 0.6279 \\ 
C11 & 4FGLJ1733.2-2915 & 17:33:13.70 & -29:15:25.6 &  3.696 & b360 & 0.5950 \\  
C12 & 4FGLJ1734.0-2933 & 17:34:03.07 & -29:33:02.9 &  3.156 & b360 & 0.6732 \\
C13 & 4FGLJ1734.5-2818 & 17:34:35.52 & -28:18:07.9 &  3.948 & b361 & 0.5351 \\ 
C14 & 4FGLJ1737.2-2421 & 17:37:13.92 & -24:21:58.7 &  3.324 & b391 & 0.4342 \\  
C15 & 4FGLJ1737.3-3332 & 17:37:22.35 & -33:32:01.3 &  3.756 & b316 & 0.9934 \\ 
C16 & 4FGLJ1738.1-2453 & 17:38:10.20 & -24:53:37.3 &  4.650 & b377 & 0.4810 \\
C17 & 4FGLJ1738.8-3418 & 17:38:53.40 & -34:18:36.0 &  4.920 & b302 & 0.8755 \\ 
C18 & 4FGLJ1739.2-2717 & 17:39:16.35 & -27:17:42.4 &  4.416 & b362 & 0.4516 \\ 
C19 & 4FGLJ1740.7-2640 & 17:40:44.57 & -26:40:37.2 &  3.426 & b362 & 0.7807 \\ 
C20 & 4FGLJ1741.1-3328 & 17:41:10.39 & -33:28:04.4 &  4.908 & b302 & 0.7213 \\ 
C21 & 4FGLJ1741.3-3357 & 17:41:22.80 & -33:57:23.4 &  4.446 & b302 & 0.5486 \\  
C22 & 4FGLJ1742.8-2246 & 17:42:53.88 & -22:46:18.5 &  4.542 & b379 & 0.2885 \\ 
C23 & 4FGLJ1743.4-2406 & 17:43:28.23 & -24:06:14.4 &  4.116 & b378 & 0.2891 \\ 
C24 & 4FGLJ1744.9-3322 & 17:44:56.07 & -33:22:19.9 &  4.494 & b303 & 0.6032 \\  
C25 & 4FGLJ1745.6-3626 & 17:45:39.34 & -36:26:17.9 &  3.744 & b273 & 0.3488 \\  
C26 & 4FGLJ1746.1-2541 & 17:46:11.54 & -25:41:17.2 &  3.240 & b349 & 0.5557 \\
C27 & 4FGLJ1747.0-3505 & 17:47:05.02 & -35:05:56.0 &  3.072 & b288 & 0.2421 \\  
C28 & 4FGLJ1747.7-2141 & 17:47:45.58 & -21:41:22.9 &  4.374 & b380 & 0.4450 \\ 
C29 & 4FGLJ1747.9-3224 & 17:47:56.28 & -32:24:45.4 &  3.390 & b303 & 1.0774 \\  
C30 & 4FGLJ1750.6-1906 & 17:50:39.46 & -19:06:36.4 &  3.402 & b396 & 0.3183 \\
C31 & 4FGLJ1750.9-3301 & 17:50:55.15 & -33:01:12.4 &  3.918 & b289 & 0.3749 \\  
C32 & 4FGLJ1752.3-2914 & 17:52:18.53 & -29:14:57.1 &  3.330 & b320 & 0.3951 \\  
C33 & 4FGLJ1752.3-3139 & 17:52:20.26 & -31:39:07.6 &  3.390 & b290 & 0.5927 \\
C34 & 4FGLJ1753.3-3325 & 17:53:23.93 & -33:25:00.8 &  4.224 & b275 & 0.2869 \\  
C35 & 4FGLJ1754.6-2933 & 17:54:39.14 & -29:33:06.5 &  3.456 & b306 & 0.2992 \\ 
C36 & 4FGLJ1754.8-3200 & 17:54:48.79 & -32:00:02.5 &  3.978 & b290 & 0.3613 \\ 
C37 & 4FGLJ1757.4-3125 & 17:57:24.94 & -31:25:07.3 &  3.792 & b291 & 0.3970 \\ 
C38 & 4FGLJ1758.0-2953 & 17:58:02.52 & -29:53:21.5 &  3.786 & b292 & 0.3889 \\ 
C39 & 4FGLJ1758.3-1920 & 17:58:23.09 & -19:20:36.2 &  4.224 & b368 & 0.7412 \\  
C40 & 4FGLJ1800.5-2910 & 18:00:30.10 & -29:10:24.6 &  4.200 & b292 & 0.2789 \\
C41 & 4FGLJ1801.0-2802 & 18:01:02.90 & -28:02:21.5 &  4.002 & b307 & 0.3531 \\  
C42 & 4FGLJ1802.1-2652 & 18:02:10.51 & -26:52:17.0 &  4.518 & b308 & 0.8211 \\
C43 & 4FGLJ1805.1-3618 & 18:05:06.36 & -36:18:21.6 &  3.114 & b232 & 0.0789 \\ 
C44 & 4FGLJ1808.4-3522 & 18:08:28.51 & -35:22:24.6 &  3.330 & b233 & 0.0759 \\  
C45 & 4FGLJ1808.5-3701 & 18:08:30.43 & -37:01:50.2 &  3.288 & b218 & 0.0520 \\
C46 & 4FGLJ1809.2-2726 & 18:09:13.82 & -27:26:44.2 &  4.218 & b280 & 0.1862 \\
C47 & 4FGLJ1811.0-2725 & 18:11:03.44 & -27:25:19.2 &  4.302 & b280 & 0.1787 \\
C48 & 4FGLJ1814.7-3420 & 18:14:45.75 & -34:20:28.3 &  3.564 & b234 & 0.0652 \\
C49 & 4FGLJ1816.4-2727 & 18:16:25.47 & -27:27:22.7 &  4.698 & b266 & 0.1222 \\ 
C50 & 4FGLJ1817.9-3334 & 18:17:55.17 & -33:34:21.7 &  3.846 & b221 & 0.0635 \\ 
C51 & 4FGLJ1819.9-2926 & 18:19:57.58 & -29:26:20.8 &  4.992 & b252 & 0.0920 \\  
C52 & 4FGLJ1820.7-3217 & 18:20:45.81 & -32:17:26.5 &  4.944 & b222 & 0.0726 \\
C53 & 4FGLJ1828.2-3252 & 18:28:13.49 & -32:52:10.6 &  4.998 & b208 & 0.0599 \\  
 \hline
\end{tabular}
\end{table*}

\subsection{Near- and mid-IR photometry}

Our main goal is to identify the selected UGS with
near- and mid-IR photometry counterparts using the VVV and WISE photometry, respectively. \cite{Pichel2020} analysed the four blazars located in the VVV region that were
identified in the Multi-frequency Catalogue of Blazars (\citealt{Massaro2015}) as counterparts to 3FGL sources.
They defined a specific region with a radius twice the positional uncertainties associated with the high-energy sources and performed a search for all infrared sources within this area. 
The photometry was conducted in the five VVV
passbands: Z, Y, J, H and K$_\mathrm{s}$ using
the combination of SExtractor (Source-Extractor) + PSFEx (PSF Extractor) \citep{Bertin2011} to assess all the sources in the region as described in \cite{Baravalle2018}.
The blazars were characterised by their near- and mid-IR properties from VVV and WISE surveys, respectively showing
different colours in the infrared diagrams. 
The photometric results of the blazar 5BZQJ1802-3940 \citep{Pichel2020} obtained with SExtractor+PSFEx were also compared in \cite{Donoso2020} with the data product provided by Cambridge Astronomical Survey Unit (CASU; \citealt{Emerson2006}). Both approaches produce comparable results and the studied blazar occupied a similar position in the colour-colour diagrams.

In this work, we analysed all the VVV sources with CASU photometry lying within the positional uncertainty region of the UGS. For this purpose, for each 4FGL sources, we defined a search area centred on the UGS, with radius defined by the semi-major axis of the ellipse (values reported in Table~\ref{tab1}-\ref{tab3}). We used the positions of the NIR sources, the object classification and the
aperture magnitudes within an aperture of radius of 3 pixels, which correspond to  $\sim$1 arcsec \citep{Minniti2010, Saito2010}. In this way all the Fermi-LAT sources were surveyed in an homogeneous way.

Regarding the mid-IR, the WISE mission observed the sky in four passbands: [3.4], [4.6], [12] and [22] $\mu$m, with an angular resolution of 6.1, 6.4, 6.5, and 12.0 arcsec achieving 5$\sigma$ point source sensitivities of 0.08, 0.11, 1, and 6 mJy, respectively, in unconfused regions on the ecliptic \citep{Wright2010}. The angular resolution in the [3.4], [4.6] and [12] $\mu$m passbands is 6 arcsec while in the [22] $\mu$m passband is 12 arcsec. All the WISE magnitudes are in the Vega system.

\subsubsection{Colour-Magnitude and Colour-Colour diagrams}

\textbf{Near-IR: VVV survey}
\label{sect:VVV diagrams}

In this section, we use NIR magnitudes and colours of all the VVV objects in the regions of the 4FGL sources. 
The magnitudes were corrected by interstellar extinction along the line-of-sight, using the dust maps of \citet{Schlafly2011} and the VVV NIR relative extinction coefficients of \cite{Catelan2011}. Then, we obtained the colours for all the sources.

\citet{Baravalle2018} defined extragalactic sources using the colour criteria  0.5 $<$ (J-K$_\mathrm{s}$) $<$ 2.0 mag;
0.0 $<$ (J-H) $<$ 1.0 mag; and 0.0 $<$ (H-K$_\mathrm{s}$) $<$ 2.0 mag with the colour constraint
(J-H) + 0.9 (H-K$_\mathrm{s}$) $>$ 0.44 mag to minimise false detections. 
The main result of this work is the VVV NIRGC, the catalogue of galaxies in part of the Southern Galactic disc. \citet{Massaro2016} examined the regions in the colour-colour diagrams 
using the J, H and K$_\mathrm{s}$ magnitudes from the 2MASS catalogue, specifically those occupied by Fermi-LAT blazars.  The infrared colours of the $\gamma$-ray blazars
cover a distinct region, clearly separated from the other extragalactic sources. Also, \citet{Cioni2013} performed an AGN selection using the VISTA Magellanic Survey (\citealt{Cioni2011}, VMC). In their Figure 2, they divided the JHK$_\mathrm{s}$ colour-colour space  into four regions. AGN with a point-like morphology occupy the A region while the AGN with a detected host galaxy dominate the B region. The C region contains reddened Magellanic sources and the D region is dominated mainly by stars and low-confidence AGN. Most of the
known AGN are found in the A and B regions. The appropriate cuts in colour have proven to be an extremely powerful tool for separating  stellar sources from extragalactic sources. These conclusions motivate us to develop a methodology for searching for $\gamma$-ray AGN candidates within the
positional uncertainty regions of the Fermi-LAT UGS. 

\begin{figure*}
\begin{centering}
  \includegraphics[width=0.35\textwidth,height=0.35\textwidth]{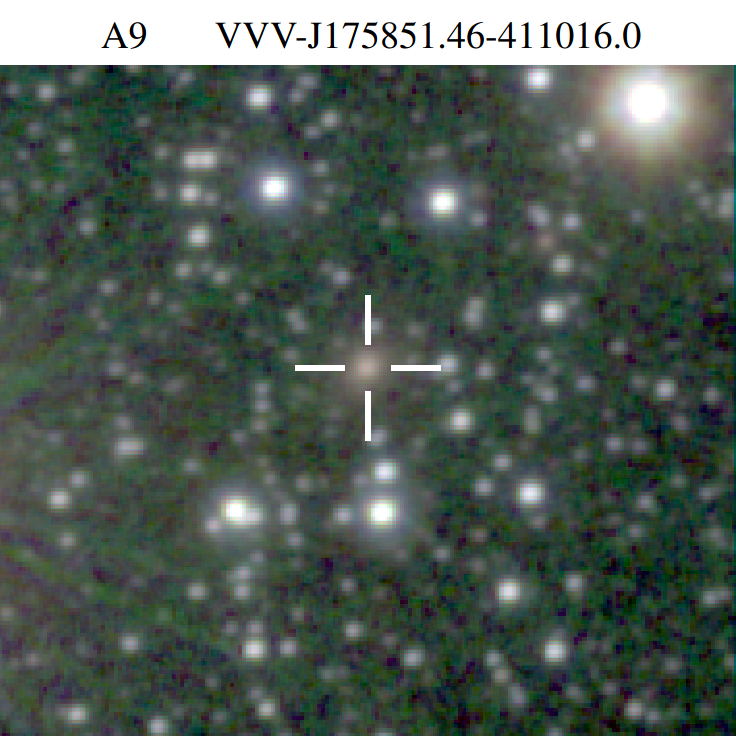}
  \includegraphics[width=0.35\textwidth,height=0.35\textwidth]{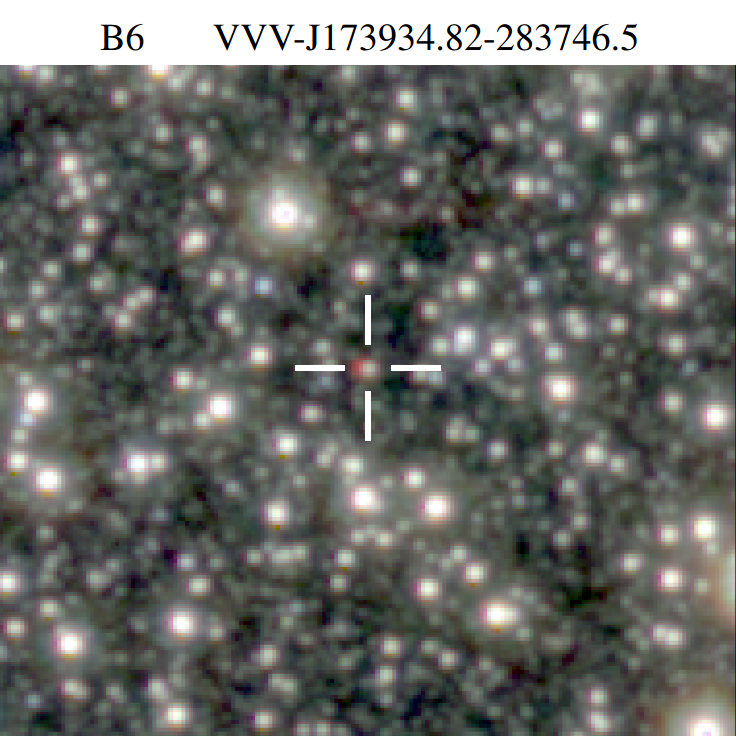}
  \includegraphics[width=0.35\textwidth,height=0.35\textwidth]{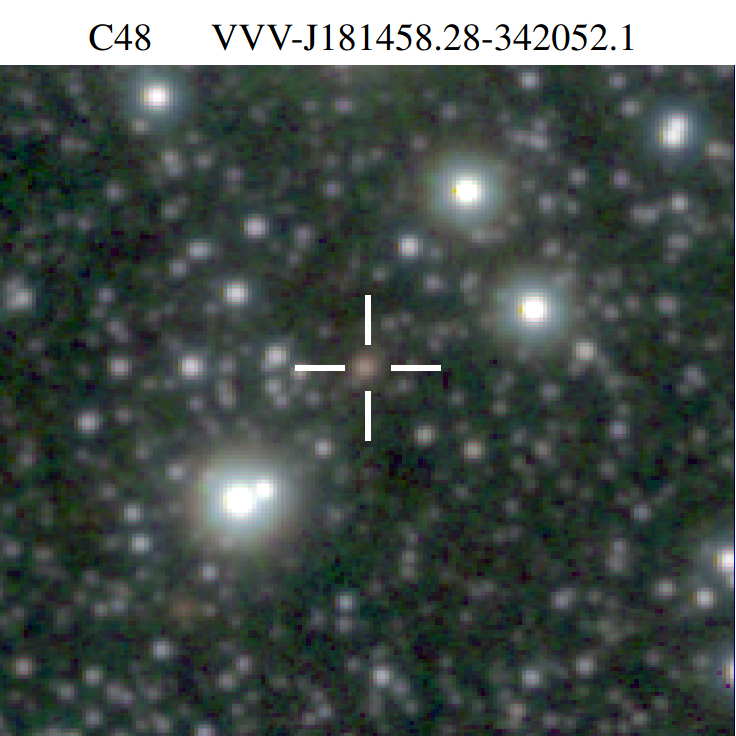}
  \includegraphics[width=0.35\textwidth,height=0.35\textwidth]{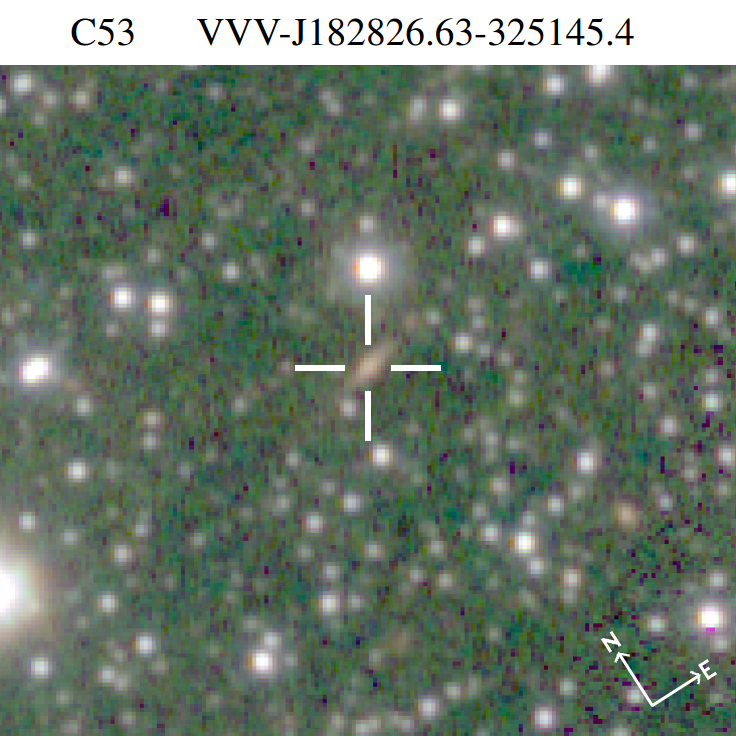}
  \caption{1$^{\prime}$ $\times$ 1$^{\prime}$ VVV colour composed images of some cases belonging to our sample of sources. 
  The orientation is shown in the bottom-right panel.}
	\label{f18}
\end{centering}
\end{figure*}

Based on the results of \cite{Baravalle2023,Massaro2016} and \cite{Cioni2011}, 
we improved the colour cuts and we selected sources that satisfy simultaneously: 0.5 < (J-K$_\mathrm{s}$) < 2.5 mag;  0.4 < (J-H) < 2.0 mag; 0.5 < (H-K$_\mathrm{s}$) < 2.0 mag and 
0.2 < (Y-J) < 2.0 mag. This selection define the possible candidates to be related to UGS. 
In addition to the colour selection, a visual inspection of the candidates in the five passbands of the survey was performed. In case of doubts, we created the false-colour red-green-blue
(RGB) images using the K$_\mathrm{s}$, H and J passbands. Figure~\ref{f18} shows some examples of these sources as 1$^{\prime}$ $\times$ 1$^{\prime}$ VVV colour composed images. 
We eliminated objects with strong contamination by bright nearby stars and those sources that have fainter K$_\mathrm{s}$ magnitudes.

\vspace{0.2cm}
{\noindent \textbf {Mid-IR: WISE}}

We applied the methodology used in \cite{Pichel2020} and \cite{DAbrusco2019} to all the Fermi-LAT sources. For the analysis, unless stated otherwise, we considered only WISE sources detected with a minimum signal-to-noise ratio of 7 in at least one passband.
Using the WGS and the WISE locus method described in \cite{DAbrusco2019}, we applied the criterion that blazars lie in a distinctive region in the 3-dimensional MIR CCD using photometry at [3.4], [4.6], [12] and [22] $\mu$m.  
The identification of WISE blazar candidates involved a selection process based on 2-dimensional projections within the CCD using the WISE locus method, as described previously. This technique may offer multiple possibilities depending on the number of identified candidates. When there is just one candidate, it is assumed to be directly associated with the Fermi-LAT source.  Nevertheless, in cases with more candidates it is difficult to determine which one is associated, making further studies essential. In addition, to improve our selection of WISE candidates, we included AGN candidates using the criteria outlined in studies by \cite{Stern2012} and \cite{Assef2018}. All identified WISE blazar candidates are also considered to be WISE AGN candidates, so all WISE candidates.

\subsubsection{Variability analysis with the VVV photometry}
\label{sec:variability}

Here, we performed the variability analysis for the objects associated to the Fermi-LAT sources. We have obtained the K$_\mathrm{s}$ passband light curves using the second version of VVV Infrared Astrometric Catalogue (VIRAC2; see \citealt{Smith2018} and Smith et al. in prep.) This is photometry based on PSF. We selected the measurements with photometric flags equal to 0 (see the catalogue) in order to obtain reliable light curves.
Their coordinates were cross-matched with the VIRAC2 assuming differences
in their positions of 1 arcsec. Twenty seven good light curves 
have a five astrometric parameter solution (a de-facto 10 epoch selection), not flagged as a probable duplicate, detected in more than 20\% of the observations that cover the source, and with a unit weight error less than 1.8. On the contrary, the
rejected objects have not met the above criteria because they are highly contaminated with nearby stars or they are too faint to have reliable magnitudes.  

In order to investigate the variability of these objects, we applied the methodology used in \cite{Pichel2020}. We examined the fractional variability amplitude, $\mathrm{\sigma_{rms}}$ \citep{Nandra1997, Edelson2002, Sandrinelli2014,Pichel2020} defined as 
$ \sigma_{rms}^2=  \frac{1}{N \mu^2}  \sum_{i=1}^{N}[ (F_{i}-\mu)^2 - \epsilon_{i}^2]$,
where N represents the number of flux values F$_\mathrm{i}$ with their  uncertainties $\epsilon_\mathrm{i}$, and $\mathrm{\mu}$ denotes the average flux. This parameter represents the excess variability that cannot be solely attributed to flux errors.
Also, we investigated the slope of the light curves, taking into consideration the results of \cite{Cioni2013}  that more than 75\% of QSO in the VCM survey exhibit a slope variation in the K$_\mathrm{s}$ passband larger than  $10^{-4}$ mag/day. 
They defined the slope of the overall K$_\mathrm{s}$ variation in the light curves that were sampled over a range of 300-600 days, 40-80 days, or shorter.
In this analysis we followed the same procedure as \citet{Baravalle2023}, we performed a linear fit of the  K$_\mathrm{s}$ light curves, considering a range of days defined by the highest and lowest variations observed in the light curve.  In all light curves, the range of days considered for this analysis varies from 1200 to $\sim$ 2300 days \citep{Baravalle2023}.

\section{Results}
\label{sec:results}

\begin{figure*}
\includegraphics[width=150mm,height=50mm]{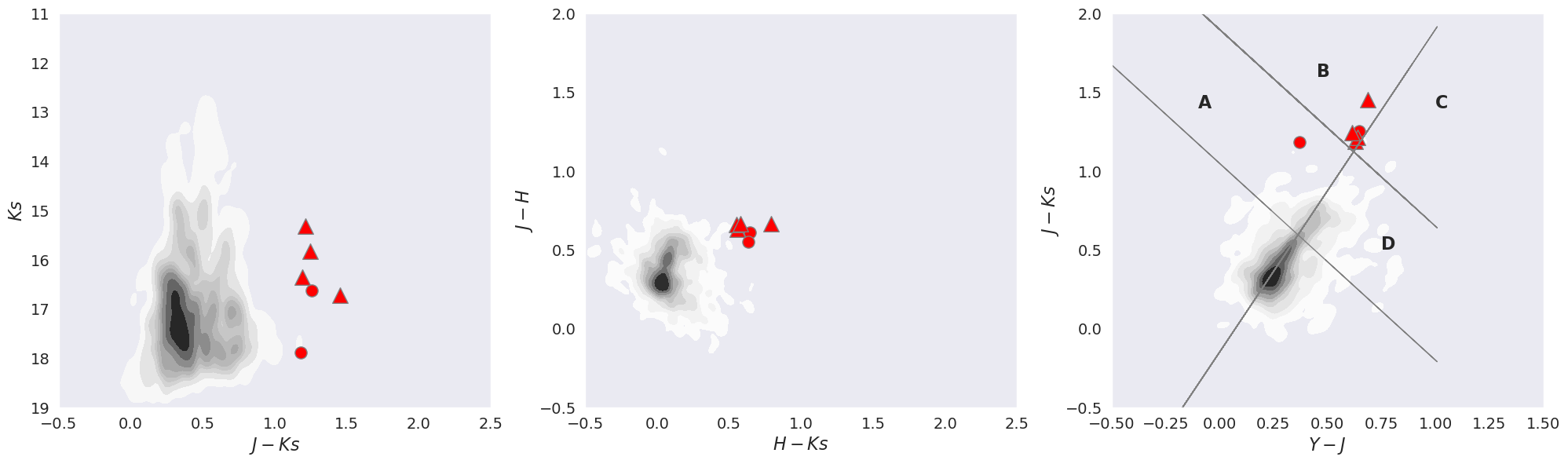}
\caption{
CMD and CCD for the field of Fermi-LAT source A9. Left, central and right panels report (J-K$_\mathrm{s}$)-K$_\mathrm{s}$ CMD, (H-K$_\mathrm{s}$)-(J-H) and (Y-J)-(J-K$_\mathrm{s}$) CCD using near-IR data from the VVV survey, respectively. The targets in red are those showing extended morphology in the images. The objects marked with circles do not have reliable variability curves; thus, the variability analysis was not performed on these targets. The objects indicated by filled triangles are those for which the variability analysis demonstrates their nature as variable sources. Grey lines defined by \citet{Cioni2013} are drawn on the YJK$_\mathrm{s}$ CCD and labels of regions defined by those authors are also indicated. Grey-scale contours correspond to density of the NIR objects, lying within the positional uncertainty region of the UGS.}   
\label{f03.vvv}
\end{figure*}

\begin{figure*}
\includegraphics[width=150mm,height=50mm]{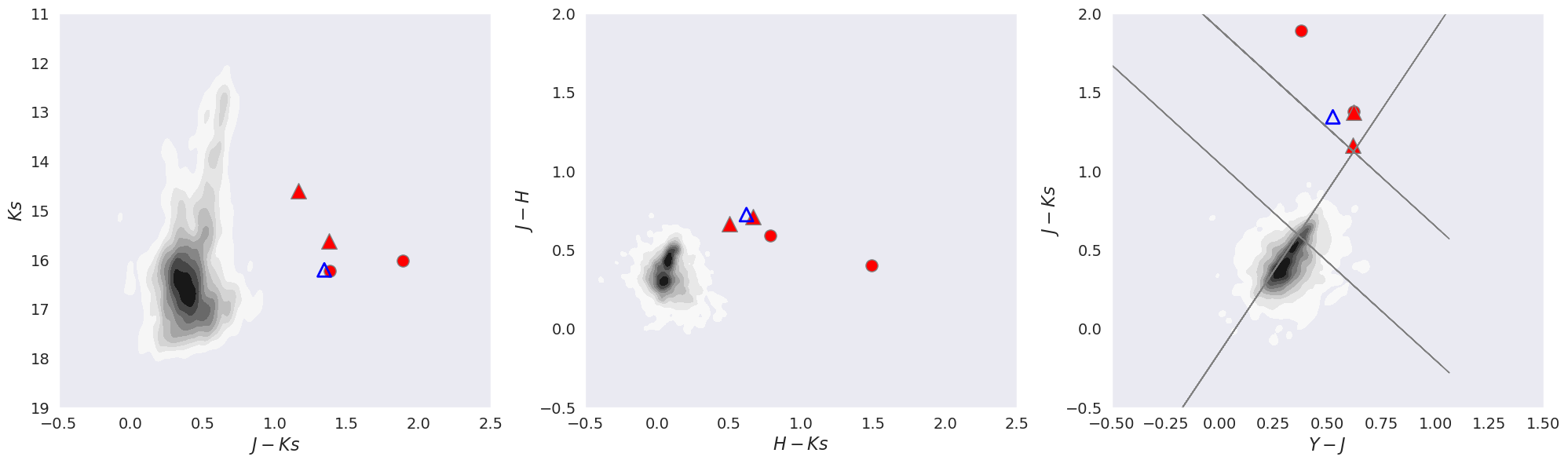}
\caption{As Fig. 4, but for Fermi source A12. Empty blue triangle represents an object with low or negligible variability and the blue colour indicates a point-like appearance.
}
\label{f04.vvv}
\end{figure*}

\begin{figure*}
\includegraphics[width=150mm,height=50mm]{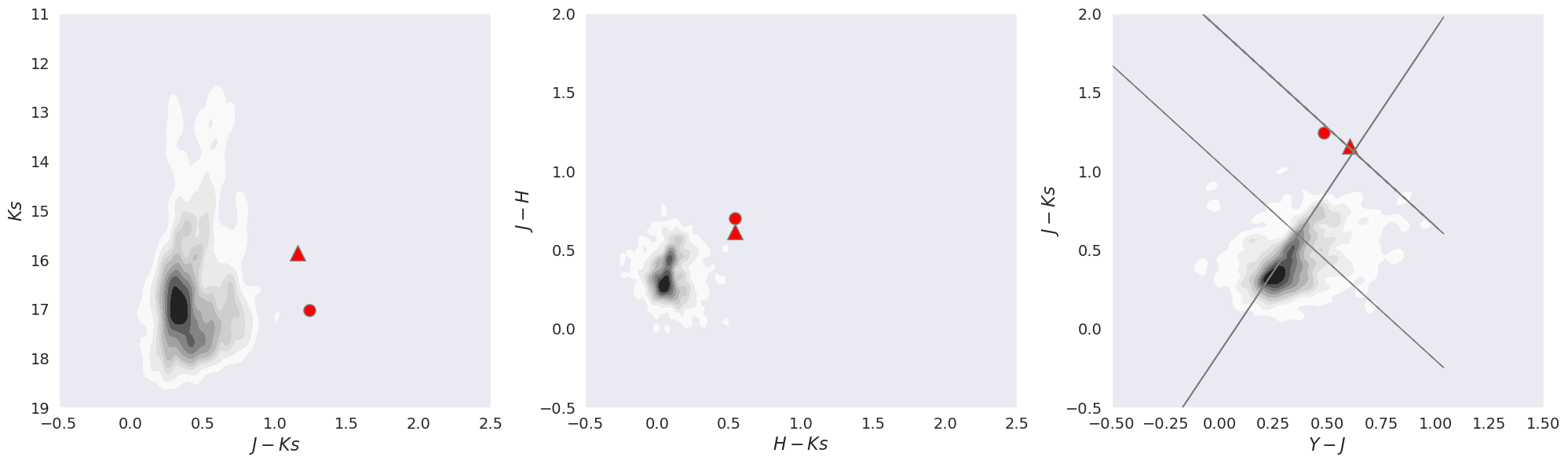}
\caption{As Fig. 4, but for Fermi source A13.}
\label{f05.vvv}
\end{figure*}

\begin{figure*}
\includegraphics[width=150mm,height=50mm]{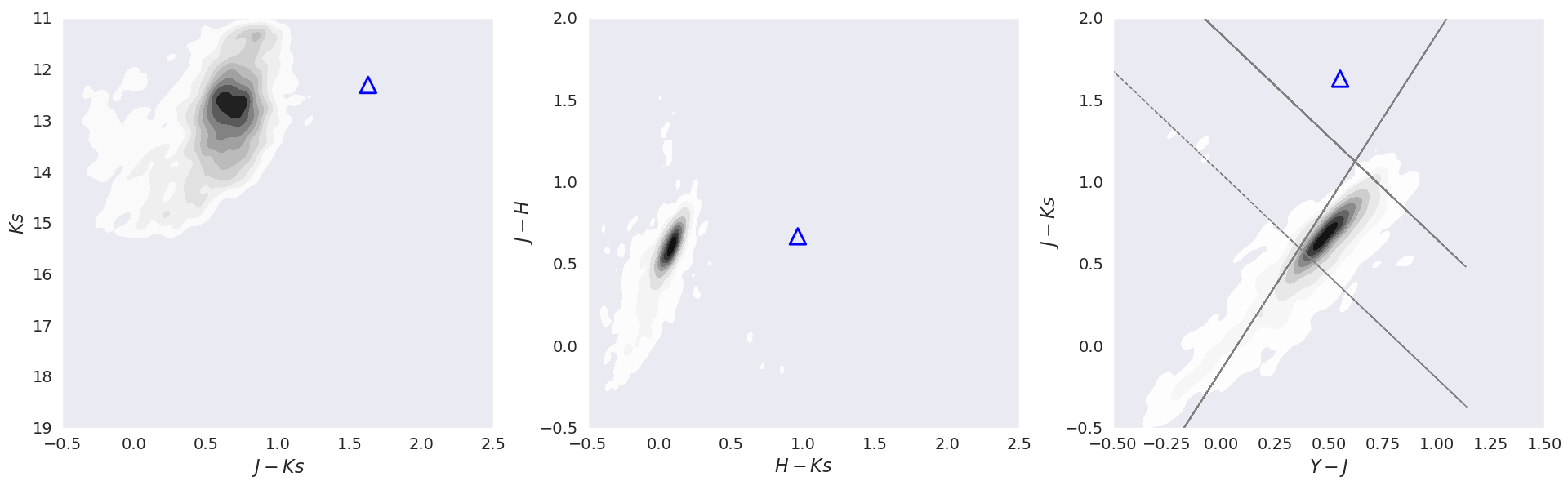}
\caption{As Fig. 4, but for Fermi source B6.  Empty blue triangle represents the candidate with low or negligible variability and the blue colour indicates a point-like morphology.}
\label{f06.vvv}
\end{figure*}

\begin{figure*}
\includegraphics[width=150mm,height=50mm]{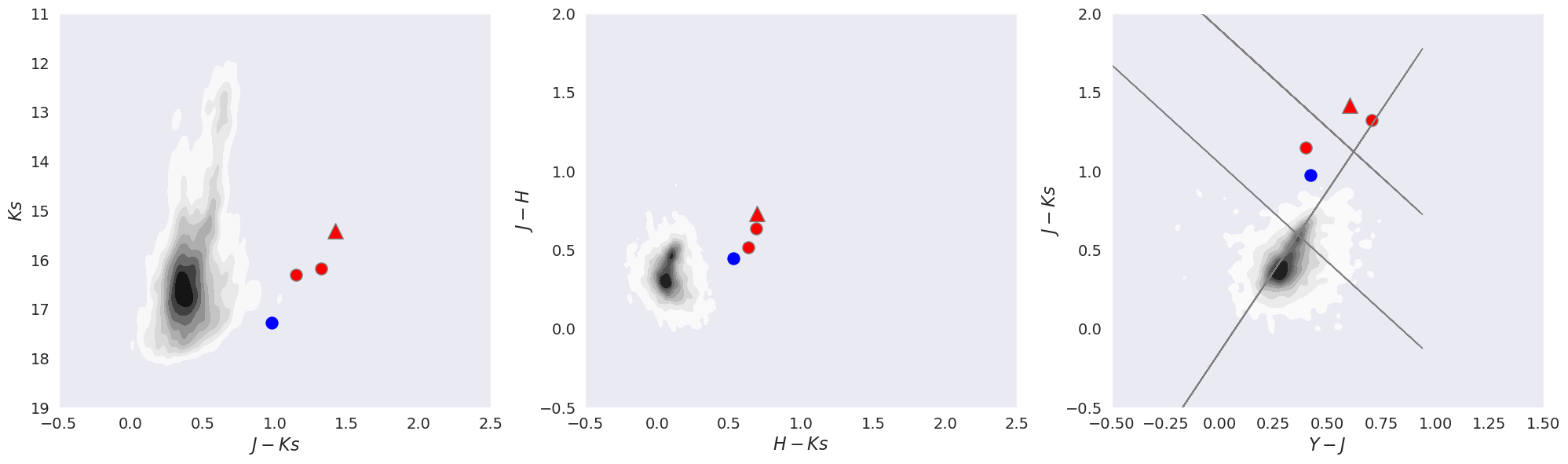}
\caption{As Fig. 4, but for Fermi source B12. The candidate with point-like morphology is indicated by blue circle. }
\label{f07.vvv}
\end{figure*}

\begin{figure*}
\includegraphics[width=150mm,height=50mm]{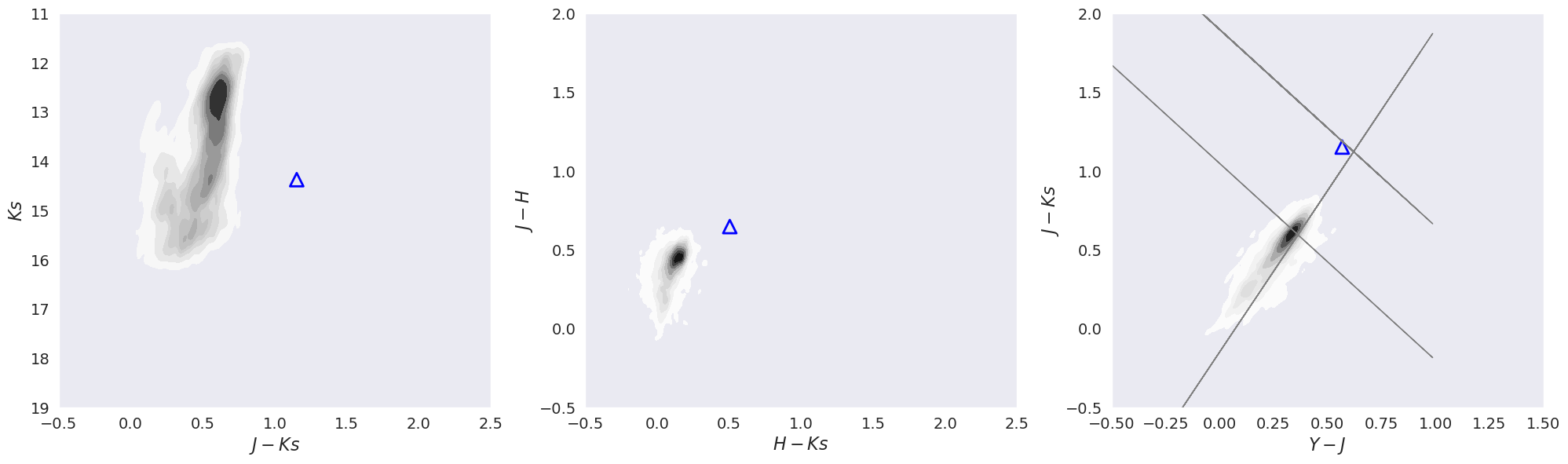}
\caption{As Fig. 4, but for Fermi source C40. Empty blue triangle represents an object with low or negligible variability and the blue colour indicates a point-like appearance.}
\label{f08.vvv}
\end{figure*}

\begin{figure*}
\includegraphics[width=150mm,height=50mm]{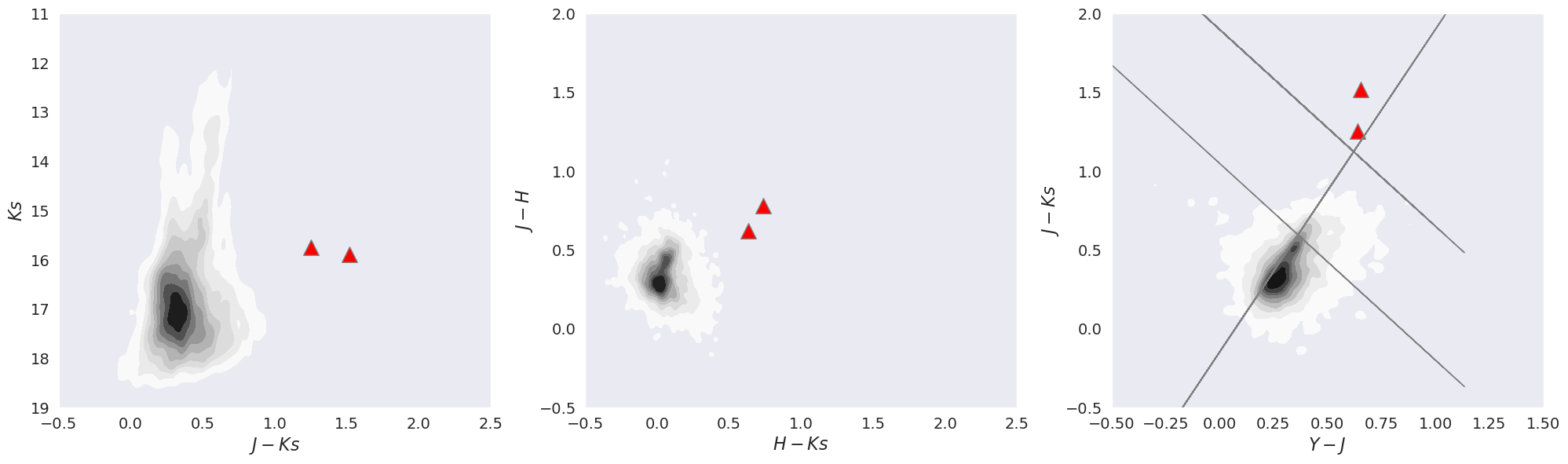}
\caption{As Fig. 4, but for Fermi source C44. }
\label{f09.vvv}
\end{figure*}

\begin{figure*}
\includegraphics[width=150mm,height=50mm]{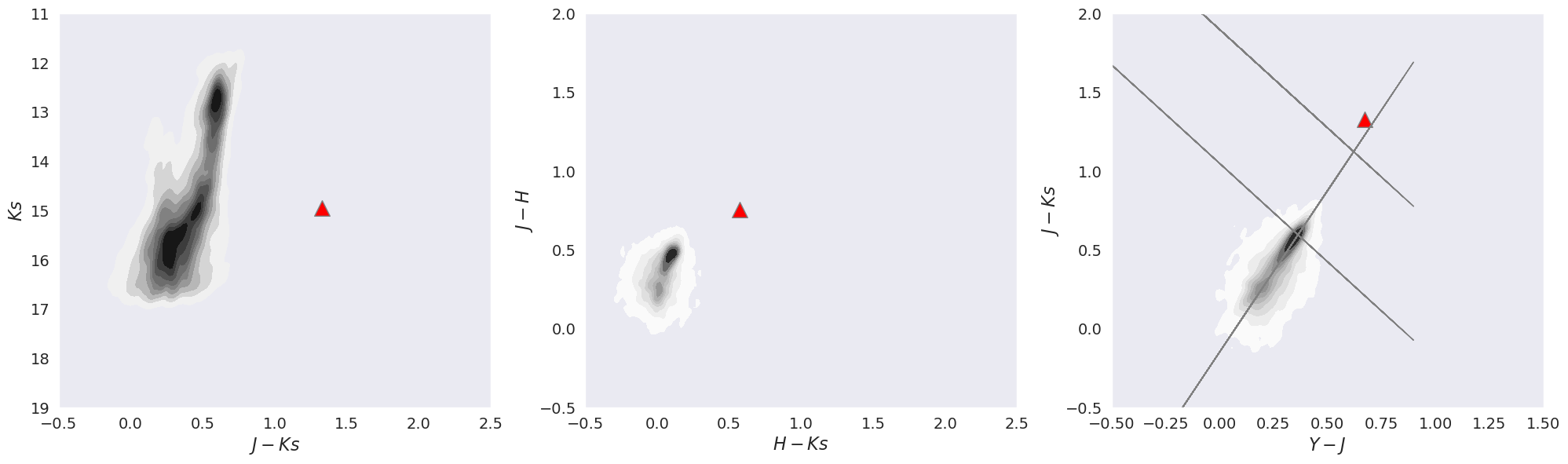}
\caption{As Fig. 4, but for Fermi source C46. }
\label{f10.vvv}
\end{figure*}

\begin{figure*}
\includegraphics[width=150mm,height=50mm]{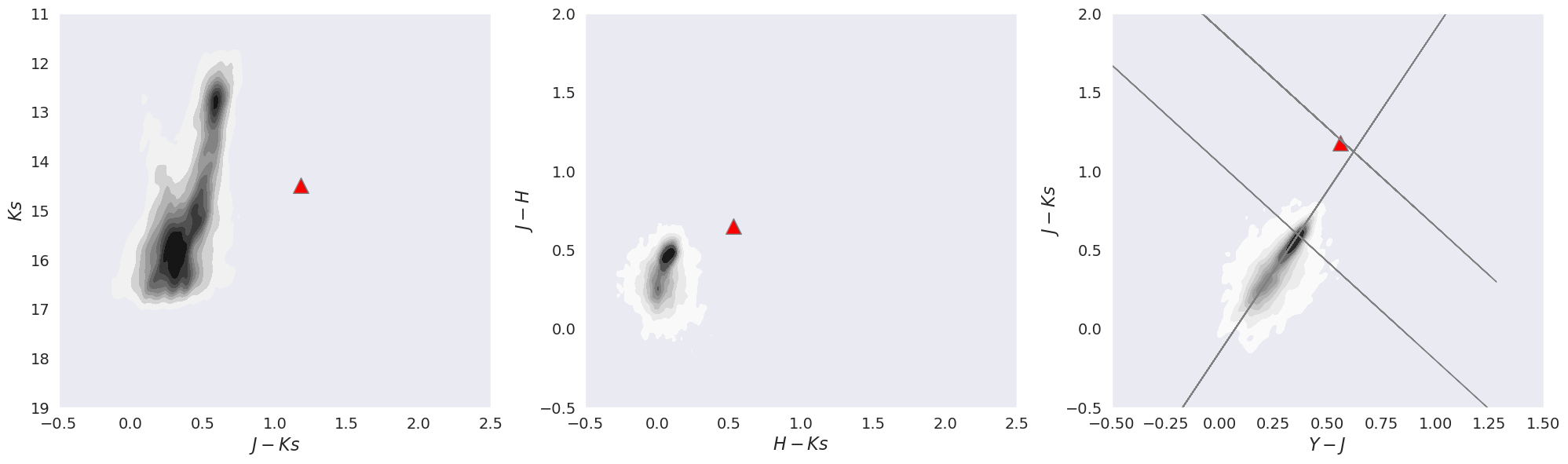}
\caption{As Fig. 4, but for Fermi source C47.}
\label{f11.vvv}
\end{figure*}

\begin{figure*}
\includegraphics[width=150mm,height=50mm]{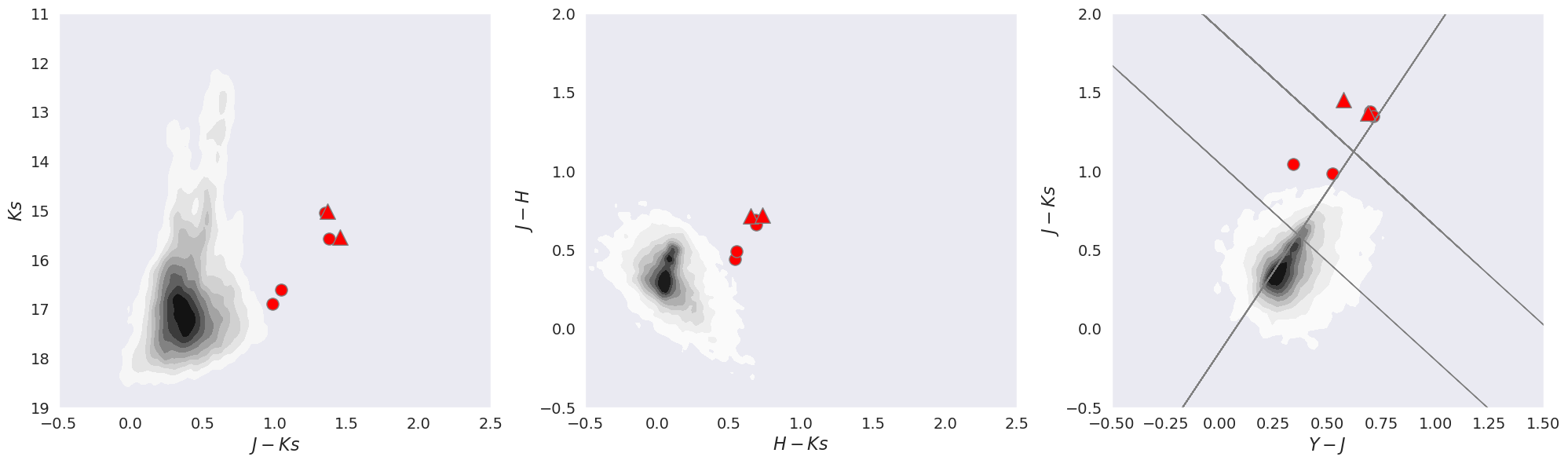}
\caption{As Fig. 4, but for Fermi source C48.}
\label{f12.vvv}
\end{figure*}

\begin{figure*}
\includegraphics[width=150mm,height=50mm]{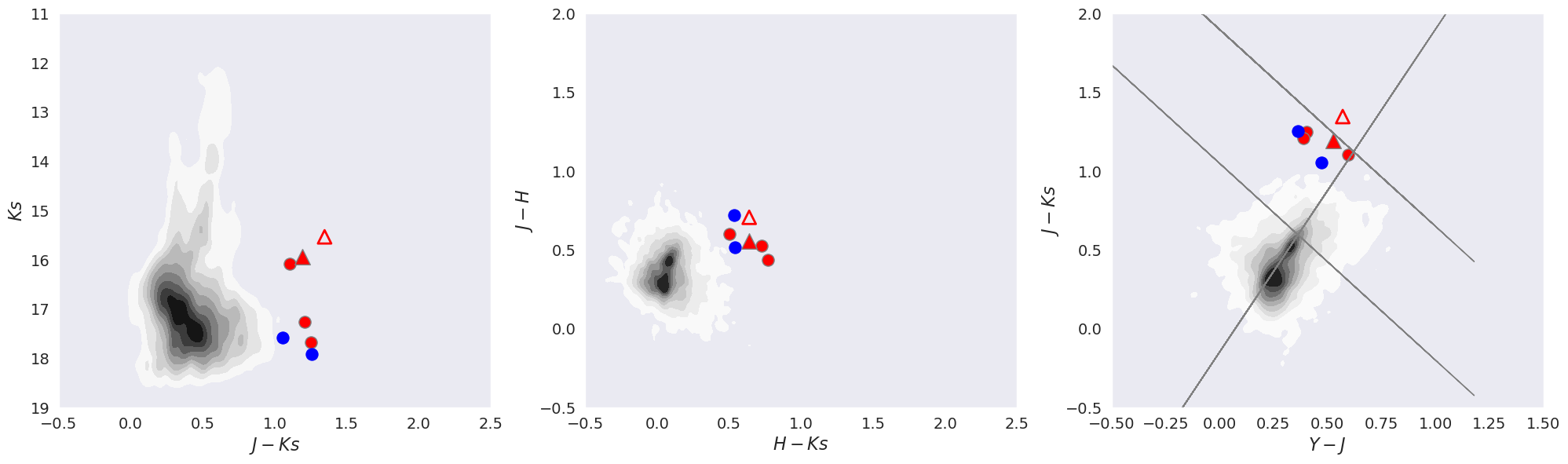}
\caption{As Fig. 4, but for Fermi source C50. Empty red triangle represents an object with low or negligible variability and the red colour indicates a extended appearance.}
\label{f13.vvv}
\end{figure*}

\begin{figure*}
\includegraphics[width=150mm,height=50mm]{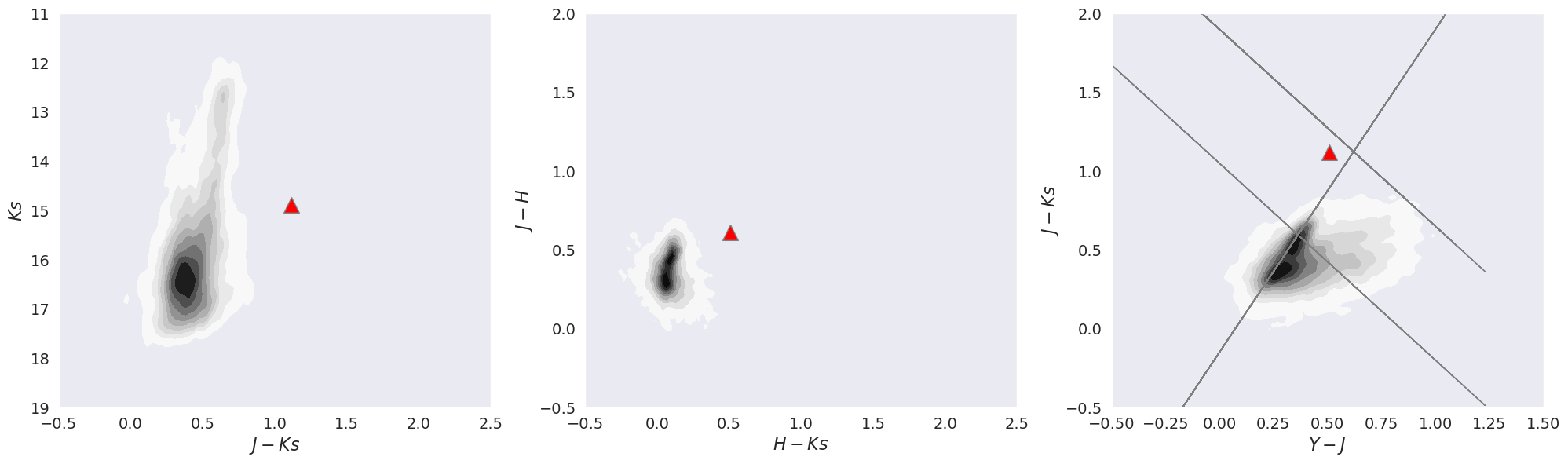}
\caption{As Fig. 4, but for Fermi source C51. }
\label{f14.vvv}
\end{figure*}

\begin{figure*}
\includegraphics[width=150mm,height=50mm]{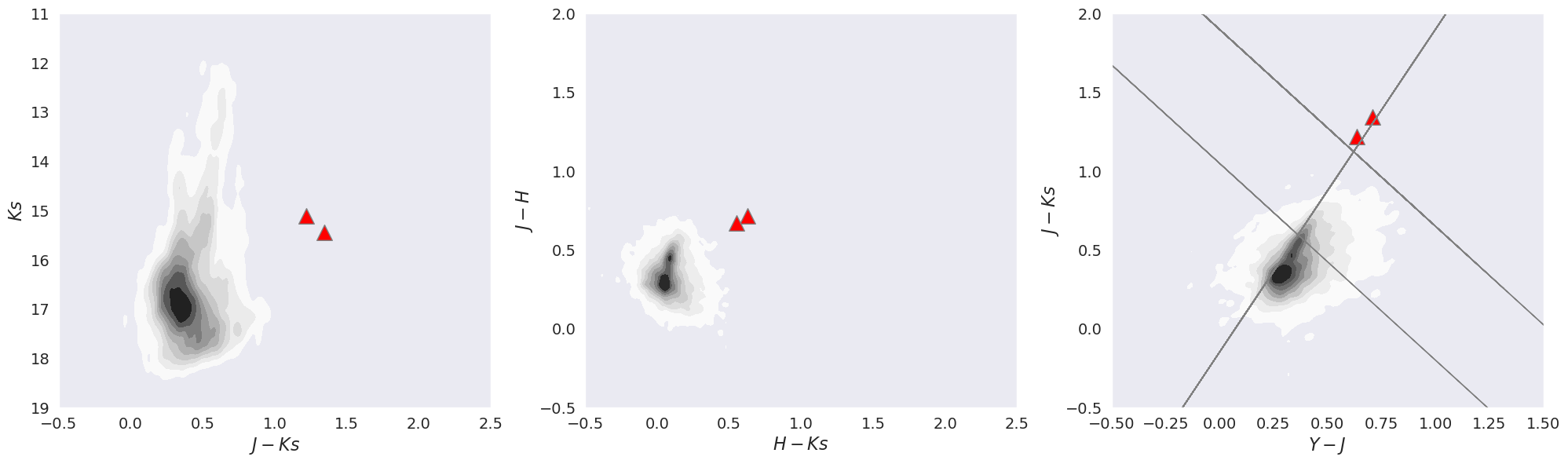}
\caption{As Fig. 4, but for Fermi source C52.  }
\label{f15.vvv}
\end{figure*}

\begin{figure*}
\includegraphics[width=150mm,height=50mm]{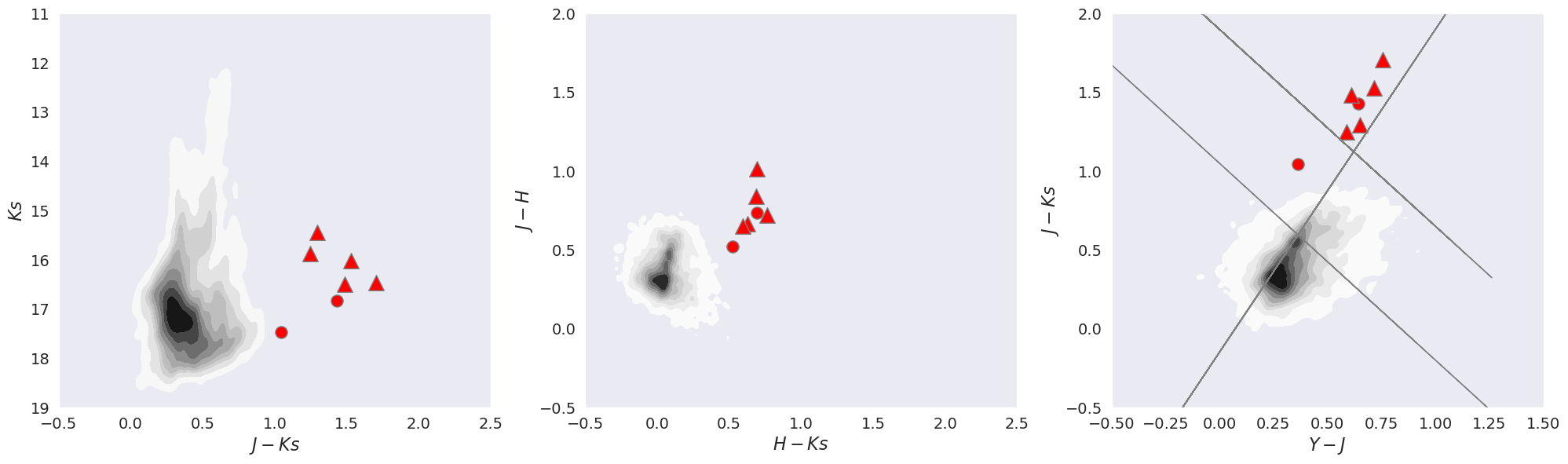}
\caption{As Fig. 4, but for Fermi source C53. }
\label{f16.vvv}
\end{figure*}

On the basis of the methodology detailed above, here, the VVV ZYJHK$_\mathrm{s}$ magnitudes, colours and K$_\mathrm{s}$ light curves of the AGN candidates are presented. As explained in subsection \ref{sect:VVV diagrams}, we have constructed colour-magnitude and colour-colour diagrams for each 4FGL.  For those 4FGL sources with candidate counterparts, the (J-K$_\mathrm{s}$)-K$_\mathrm{s}$, colour-magnitude diagram and (H-K$_\mathrm{s}$)-(J-H) and (Y-J)-(J-K$_\mathrm{s}$) colour-colour diagrams are shown in the Figures~\ref{f03.vvv} to \ref{f16.vvv}. There, grey-scale contours correspond to density of all the CASU objects found in 4FGL regions with size defined by the positional uncertainty of the Fermi-LAT source, including stellar and extragalactic sources. The regions preferentially populated for AGN candidates are: 0.5 < (J-K$_\mathrm{s}$) < 2.5 mag; 0.5 < (H-K$_\mathrm{s}$) < 2.0 mag; 0.4 < (J-H) < 2.0 mag and 0.2 < (Y-J) < 2.0 mag. The candidates were highlighted and represented by red circles for extended sources and as blue circles for objects with point-like morphology. Those AGN candidates that present variability are indicated by triangles, the colour depending on the origin of the sources:  red for galaxy-like sources and blue for stellar-like objects. Full triangles are objects that have slope in K$_\mathrm{s}$ passband higher than $10^{-4}$ mag/day and empty triangles have slopes lower than this value. Also, the regions limited with lines as defined by \citet{Cioni2013} are shown.

After careful visual inspection, we eliminated faint and contaminated sources, leaving only those that were considered VVV candidates. Thus, 7 Fermi-LAT sources have only one VVV candidate: A13, B6, B12, C40, C46, C47 and C51. Some UGS have more than one VVV candidate: the Fermi-LAT source C53 presents 5 candidates; A9, 4 candidates; A12, 3 candidates; and C44, C48, C50 and C52, 2 candidates each one. These VVV candidates are not  located in the Southern disc and therefore, there are no sources in common with the VVV NIRGC.

In Figure~\ref{fig:lightcurves}, we present the differential K$_\mathrm{s}$ light curves of the VVV sources. These curves represent the K$_\mathrm{s}$ magnitudes with the median subtracted, sampled over a period covering more than 2500 days. We noted that the overall shape of light curves is irregular, lacking any discernible periodic pattern. In some cases, we observe prominent fluctuations in brightness that resemble peaks, exhibiting statistical significance well above the value of the associated uncertainties. Table \ref{tab:variability} presents the main results of the K$_\mathrm{s}$ variability of these sources, showing the mean magnitude, $\mathrm{\sigma_{rms}}$ and the slope of the linear fits with the range of days used. Also some comments of the visual inspection of the objects are included. Most of them are early-type galaxies or the bulges of galaxies, because the near-infrared is sensitive to detecting the oldest stellar population in the galaxy. We did not include in the analysis those objects with strong crowding contamination or faint magnitudes as mentioned above.  In general, most of the studied objects exhibit moderate variability, characterised by $\mathrm{\sigma_{rms}}$ values ranging from 12.5 to 32.1. These results are in agreement with previous studies on type-1 AGN, such as those by \citet{Nandra1997, Edelson2002, Baravalle2023}. However, these values are lower than those reported for blazars \citep[e.g.,][]{Sandrinelli2014, Pichel2020}. Since type-1 AGN typically present lower variability amplitudes than blazars \citep[e.g.,][]{Ulrich1997, Mao2021}, our results suggest that these objects are potential type-1 AGN, such as quasars or Seyfert 1 galaxies. Moreover, the observed light curve slopes are $\geq$ $10^{-4}$ mag/day, comfortably lying within the limit established by \citet{Cioni2013} for quasars. On the other hand, there are four objects that present negligible variability, with very low values of $\mathrm{\sigma_{rms}}$. These objects are 
VVV-J181300.69-314505.6, VVV-J173934.82-283746.5, VVV-J180027.63-291007.4 and VVV-J181803.69-333215.7 in the regions of the Fermi-LAT sources A12, B6, C40 and C50, respectively (see Fig. \ref{fig:lightcurves} and Table \ref{tab:variability}). As expected, these objects also exhibit significantly lower slope values, typically below $10^{-5}$ mag/day. Although luminosity variability is a common feature of active galactic nuclei, the absence of variability does not necessarily rule out the possibility of an object being an AGN. It is important to note that not all AGN exhibit the same degree of variability, and certain AGN may display very low or nearly negligible levels of variability \citep[e.g.,][]{Ilic2017,Li2022,Pennock2022}. Beyond this, more than 85\% of the objects studied here show a moderate variability, and as mentioned above, these results suggest that these sources are type-1 AGN candidates. It has to be noted that this analysis is based only with photometric data. A spectroscopic study is necessary in order to investigate the nature and type of AGN.

We also searched for WISE candidates coincident with the position of the VVV candidates found before. We could not get any match between the VVV and WISE candidates with the exception of the source VVV-J173934.82-283746.5 in the Fermi-LAT B6 region. This object has a source at an angular distance of 0.64 arcsec classified as the OH/IR star 359.54+01.29 \citep{Sevenster1997}.  

\begin{figure*}
\includegraphics[width=150mm,height=50mm]{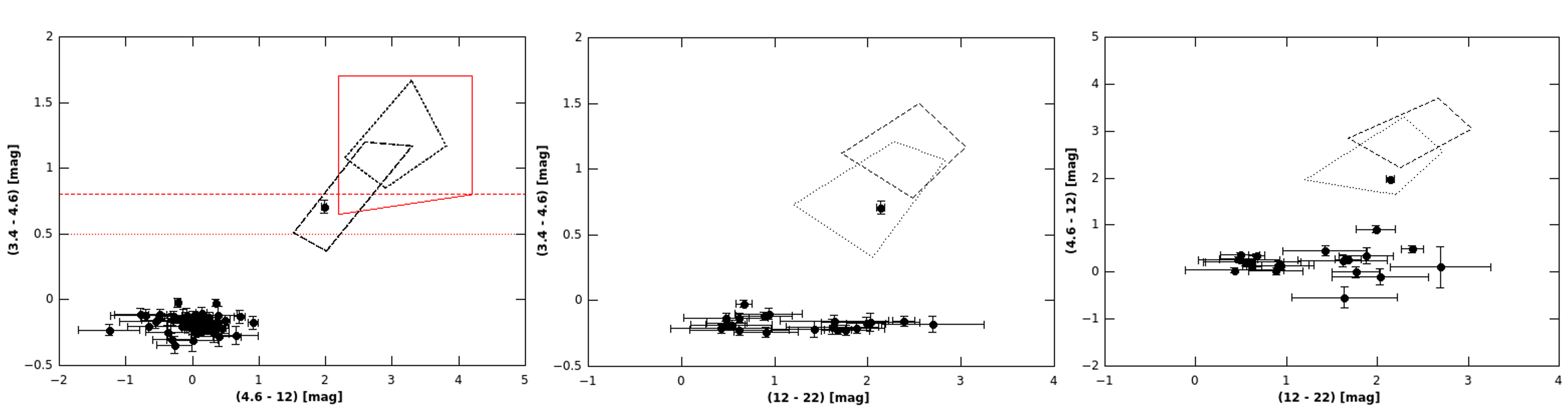}
\includegraphics[width=150mm,height=50mm]{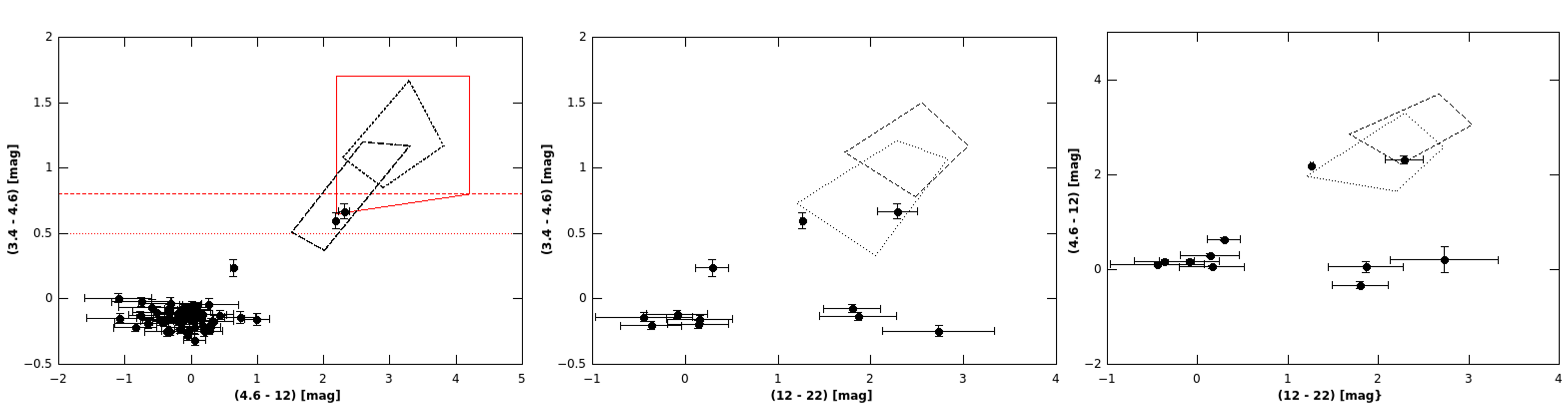}
\caption{Mid-IR colour-colour diagrams for the Fermi-LAT sources A8 (top) and B10 (bottom) using WISE  data (black dots). The two blazar classes of BZB (BL Lac) and BZQ (FSRQ) are shown in dash- and dot- black lines, respectively. The dotted and dashed red horizontal lines represent the limits for AGN from \citet{Stern2012} and \citet{Assef2018}, respectively. The solid red box denotes the defined region of QSO/AGN from \citet{Jarrett2011}. }
\label{f18.wise}
\end{figure*}

The WISE results are not as clear as those in \cite{Pichel2020} and \cite{Baravalle2023}.  
All 4 sources explored in \cite{Pichel2020} had VVV candidate counterparts, but only two of them had WISE ones. In \cite{Baravalle2023}, the 4 active galaxies had VVV and WISE counterparts.
The main difference between these two studies and the present one is that the VVV candidates were brighter in the K$_\mathrm{s}$ passband. Here, all the VVV candidates
are in the range of 14.5 to 18 mag with the exception of the candidate in the Fermi-LAT B12 region.  Another difference is the high interstellar extinction towards the fields studied here and in some cases, strong stellar contamination. In the mid-IR, the results here are more noisy in general.
Based on these results, we present candidates in the Fermi-LAT source regions both in the NIR and MIR using VVV and WISE surveys, respectively. 
However, inside the Fermi-LAT source A8 appears a WISE source (J173612.07-342204.7) that satisfies all the criteria to be a blazar candidate using the WGS method. This region has a high interstellar extinction (A$_\mathrm{Ks} = 0.9297$ mag) and the NIR CMD shows bright magnitudes without candidates to counterparts. Also, for the Fermi-LAT source B10, two WISE AGN/blazar candidates were found (J180835.96-335752.2 and J180825.25-335615.1) using the WGS  and \cite{Assef2018} methods for AGN. In Figure \ref{f18.wise} both Fermi-LAT sources are shown, using the mid-IR [3.4]-[4.6] vs [4.6]-[12], [3.4]-[4.6] vs [12]-[22] and [4.6]-[12] vs [12]-[22] colour-colour diagrams of all the WISE sources (represented
with black dots). The red box  indicates the limits used to identify QSO and Seyfert galaxies \citep{Jarrett2011}. These WISE candidates do not have VVV counterparts; thus, no other analysis and cross-match can be done in this paper. Further analysis with IR spectroscopy is needed in order to establish the nature of the WISE sources.

We might note that most of the VVV candidates are found in the B region of the colour-colour diagram defined by \citet{Cioni2013}.
Our sample of VVV candidates are centered at the position (0.6; 0.7) in the CCD (J-H) vs (H-K$_\mathrm{s}$) according to \citet{Chen2005}. For the 27 candidates listed in Table~\ref{tab:variability}, we then searched for the closest object in a circle of 30 arcsec radius 
using the SIMBAD database\footnote {https://cds.unistra.fr/} and we have not found any catalogued source, with the exception of the object in the region of the Fermi-Lat source B6 mentioned above. There have been no previous photometry or spectroscopy studies performed in these regions.    

\cite{Lefaucheur2017} obtained a sample of 595 blazar candidates from the unassociated sources within the 3FGL catalogue \citep{Acero2015}. 
They proceeded to train multivariate classifiers on samples derived from the Fermi-LAT catalogue, carefully selecting discriminant parameters. 
Within their blazar candidates, there are 30 objects in the region of the VVV survey, of which A5, B3, B4, C10, C27 and C49 in our subsamples. They classified the Fermi-LAT source A5 as BL Lac, however, there are no VVV candidates in this region because of the high interstellar extinction (A$_\mathrm{Ks}$ = 0.9213 mag).
The Fermi-LAT C10 was also classified as BL Lac and the near-IR CMD and CCD show that there are a point-like and a galaxy-like objects in the region that we have defined as possible VVV candidates. This is a region of strong crowding contamination and the K$_\mathrm{s}$ light curves of these two objects were noisy and did not satisfy our criteria. For these reasons there are no other VVV nor WISE candidates in common with these authors.

\begin{figure*}
\begin{center}
\includegraphics[width=0.30\textwidth,height=0.25\textwidth]{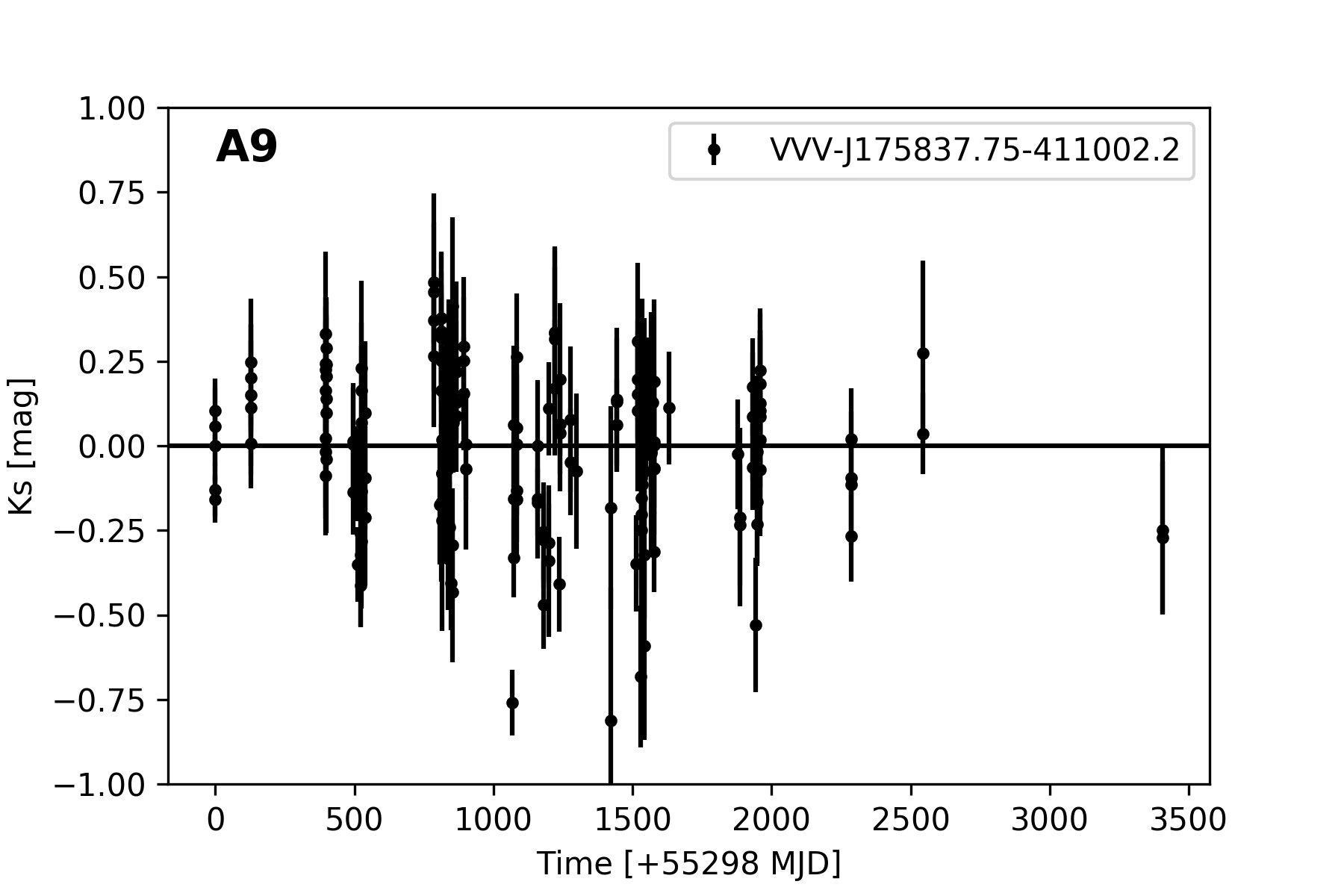}
\includegraphics[width=0.30\textwidth,height=0.25\textwidth]{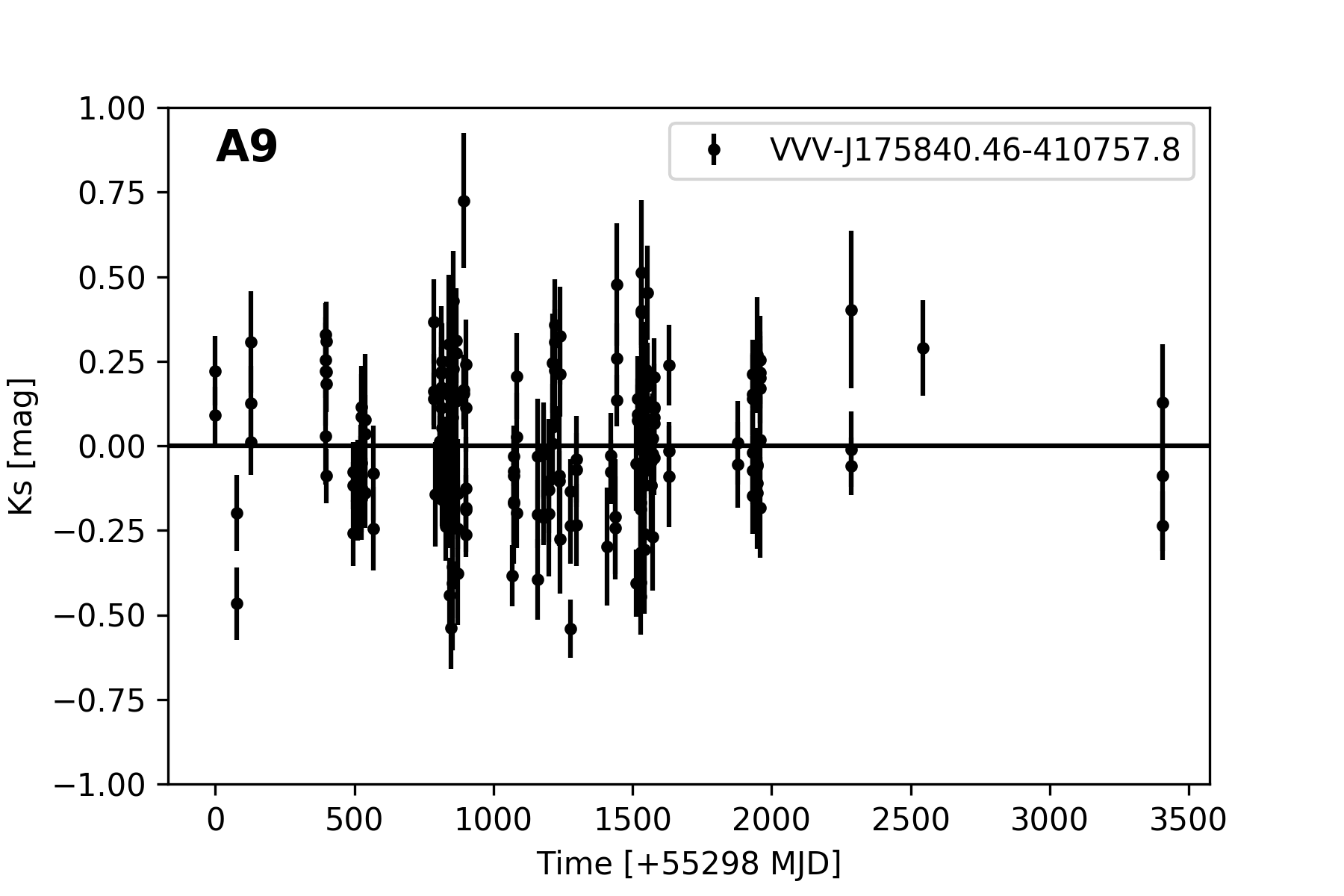}
\includegraphics[width=0.30\textwidth,height=0.25\textwidth]{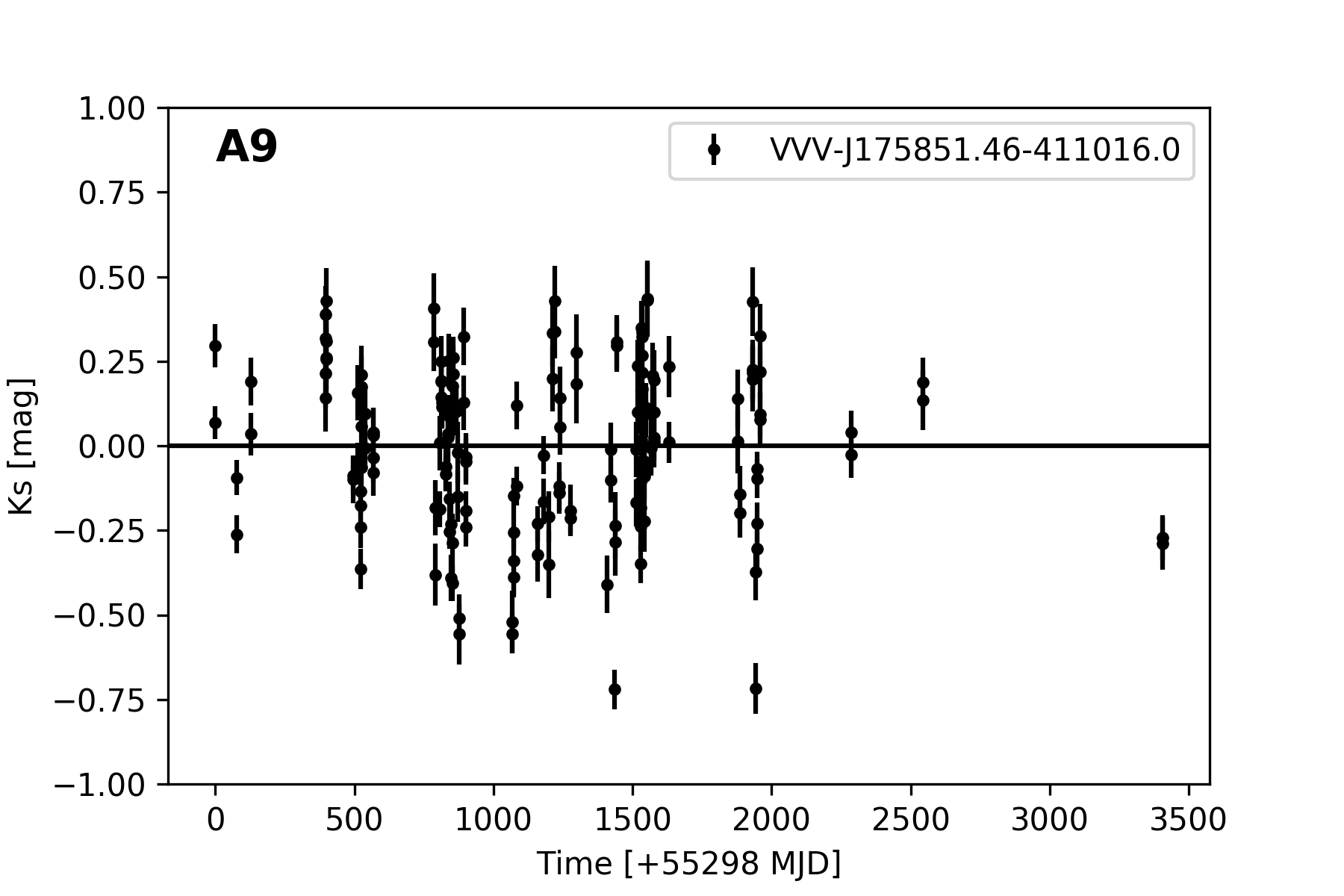}
\includegraphics[width=0.30\textwidth,height=0.25\textwidth]{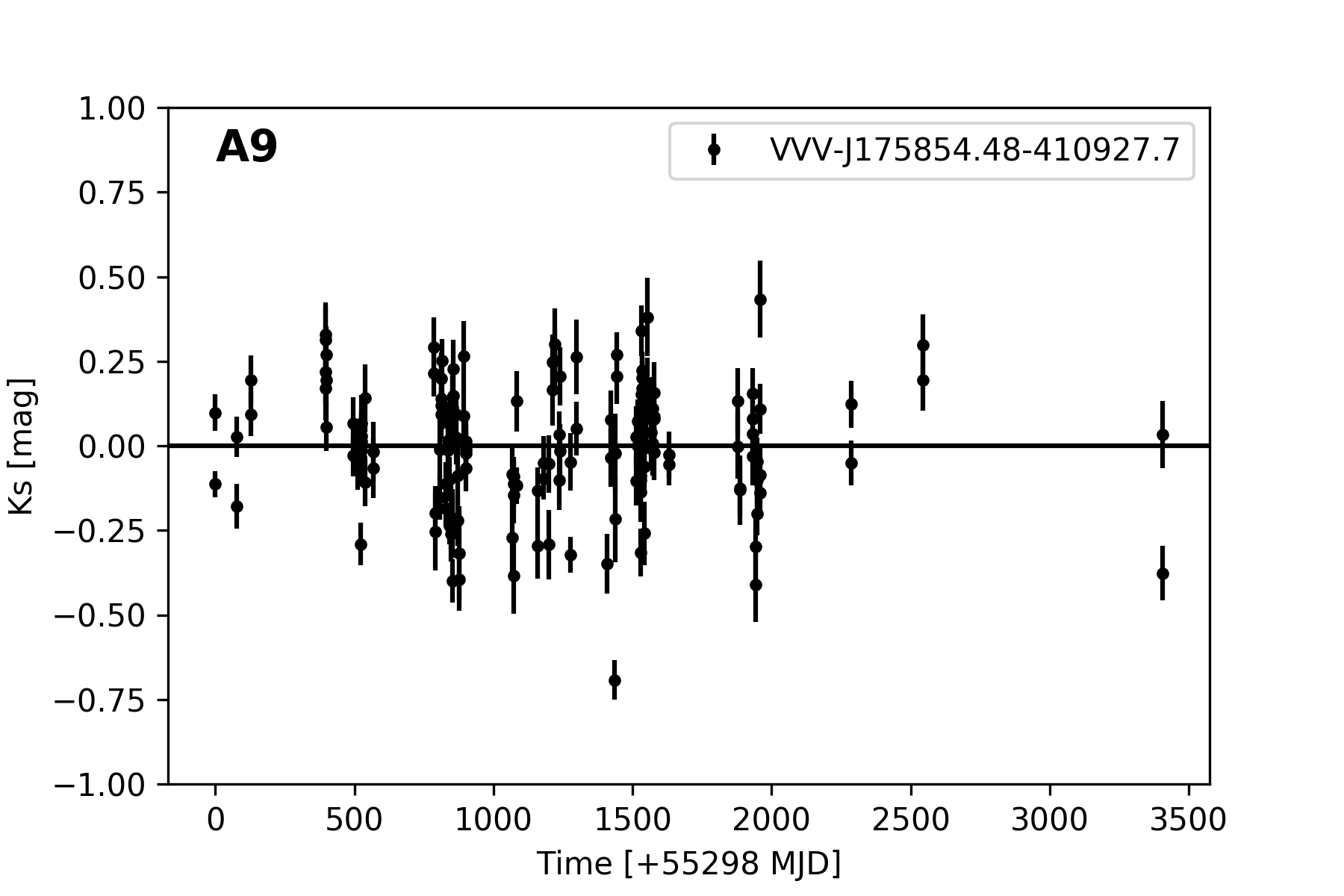}
\includegraphics[width=0.30\textwidth,height=0.25\textwidth]{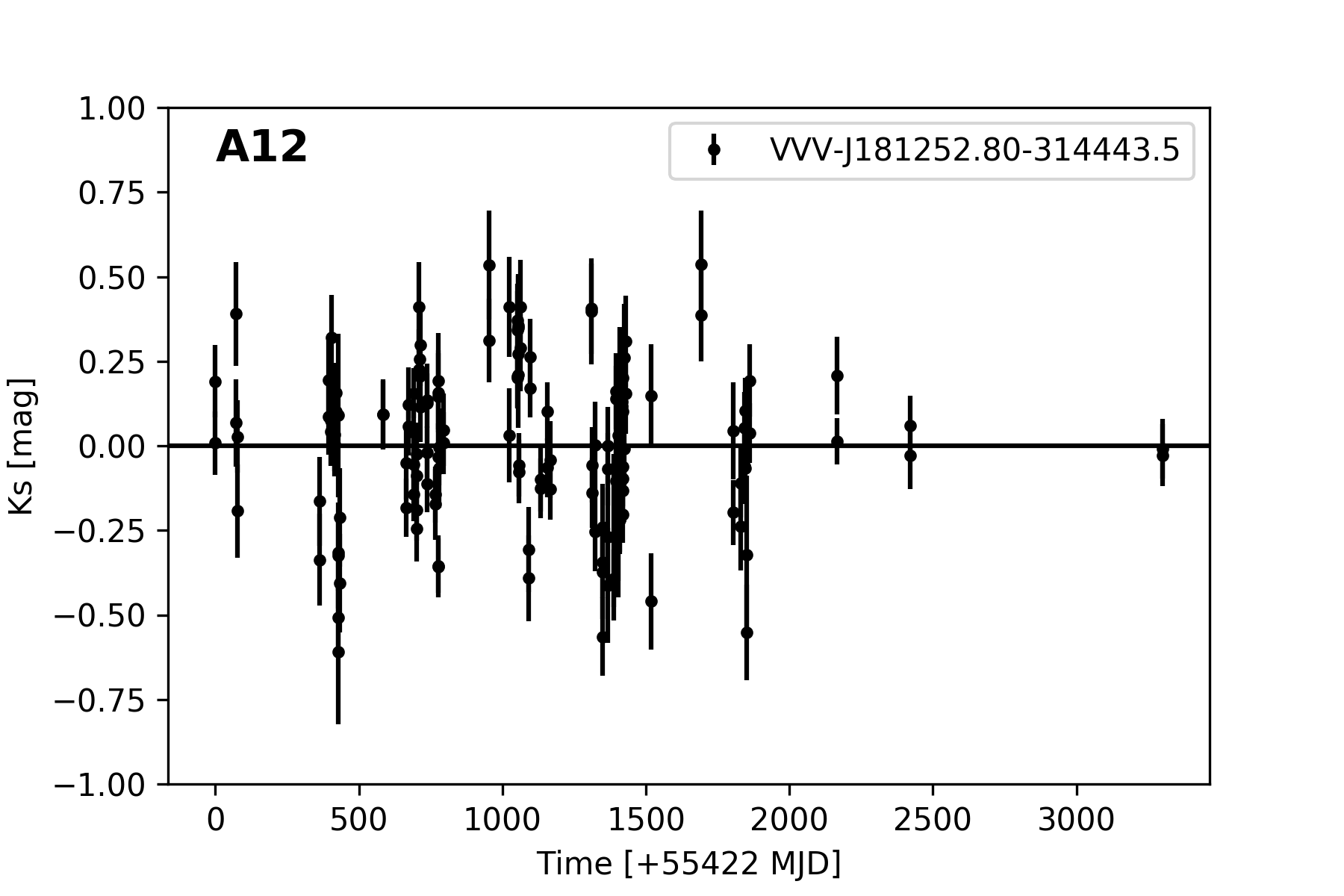}
\includegraphics[width=0.30\textwidth,height=0.25\textwidth]{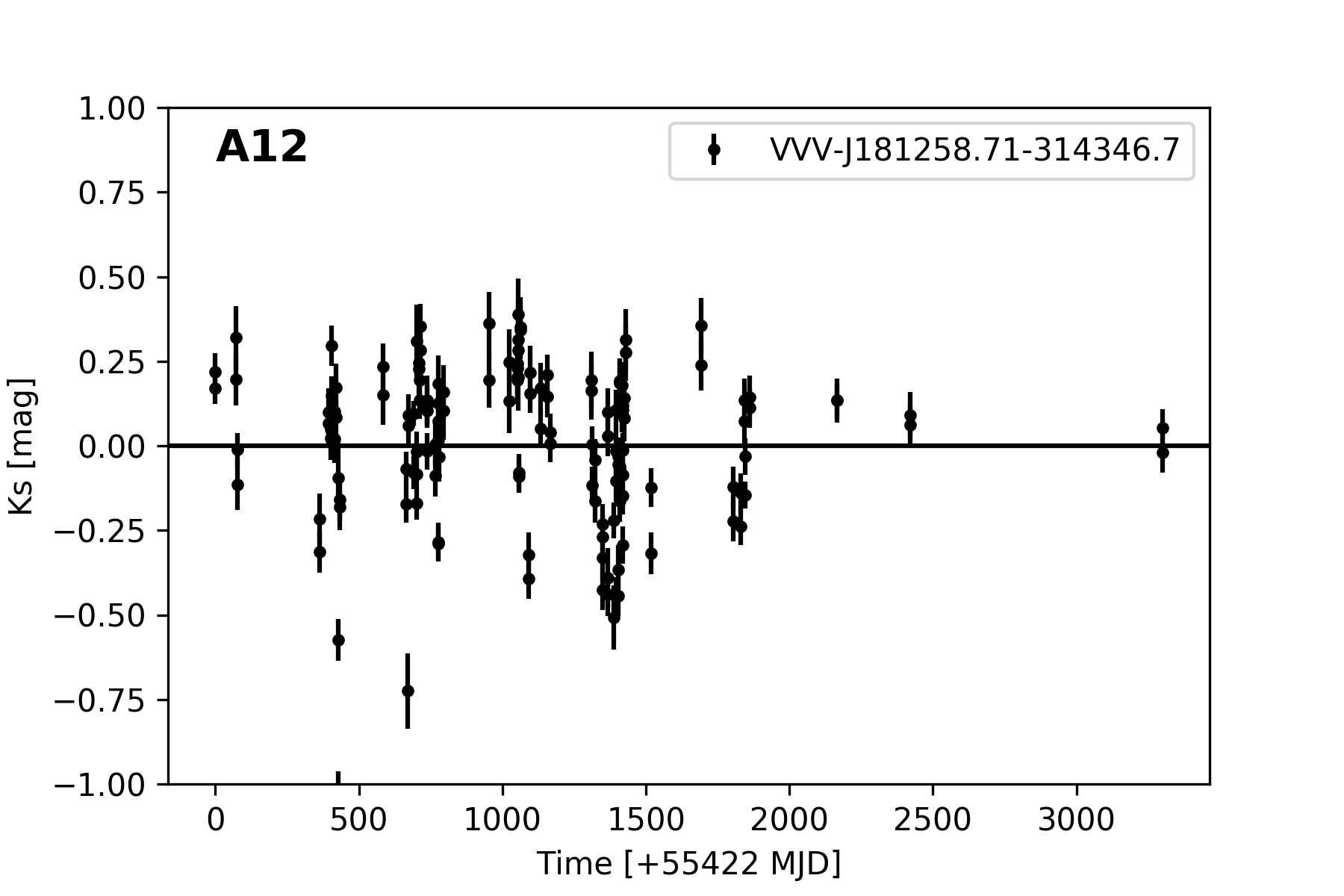}
\includegraphics[width=0.30\textwidth,height=0.25\textwidth]{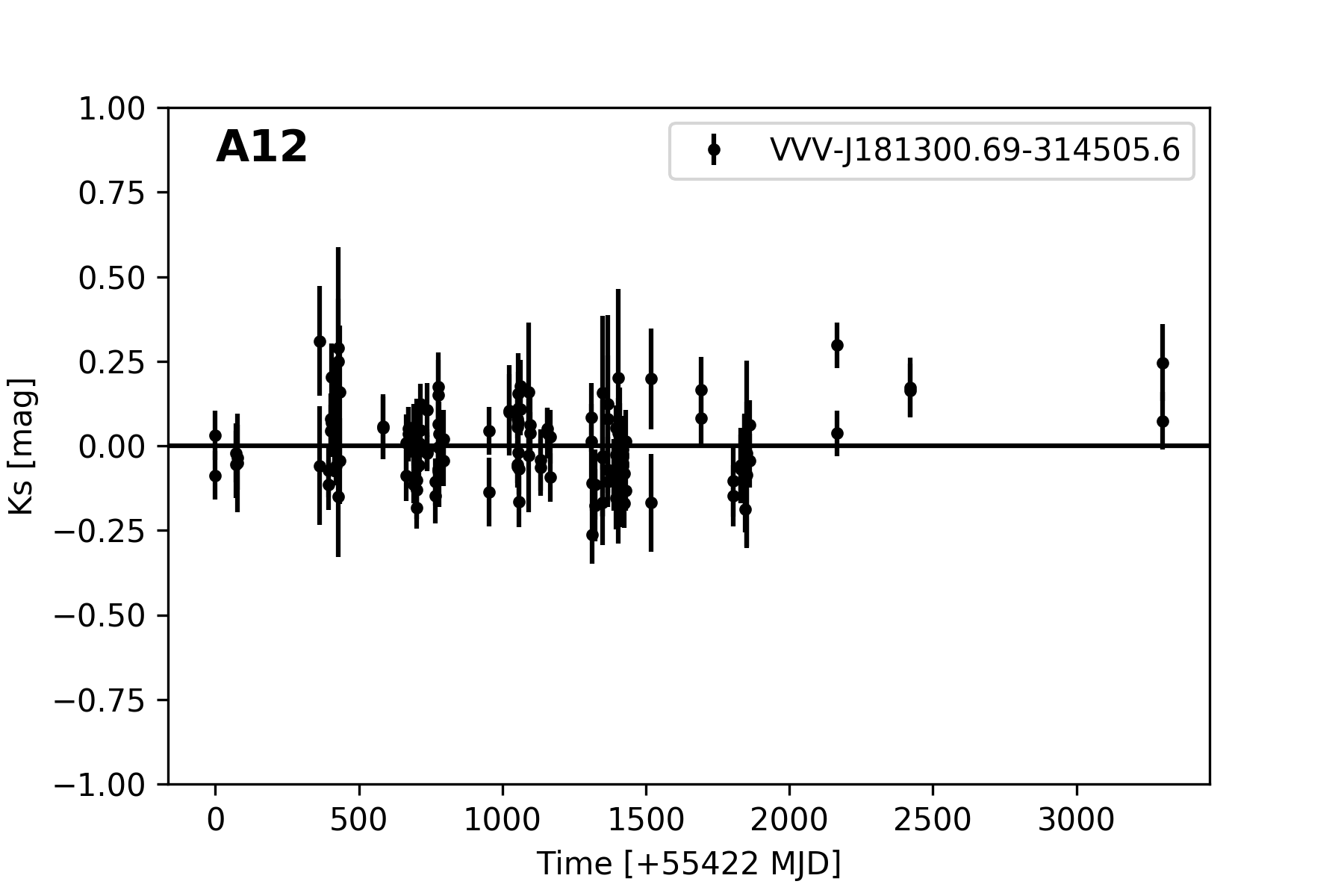}
\includegraphics[width=0.30\textwidth,height=0.25\textwidth]{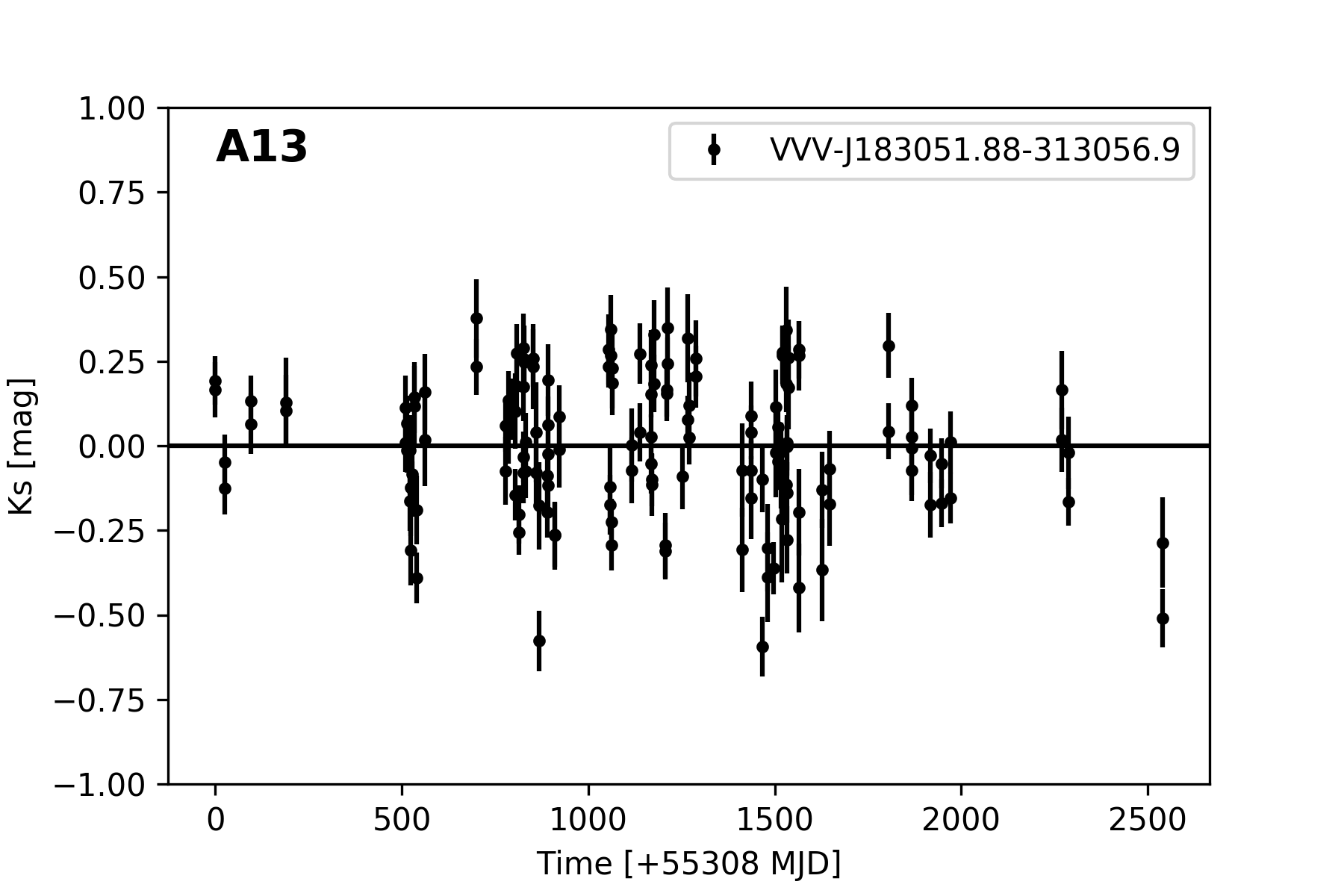}
\includegraphics[width=0.30\textwidth,height=0.25\textwidth]{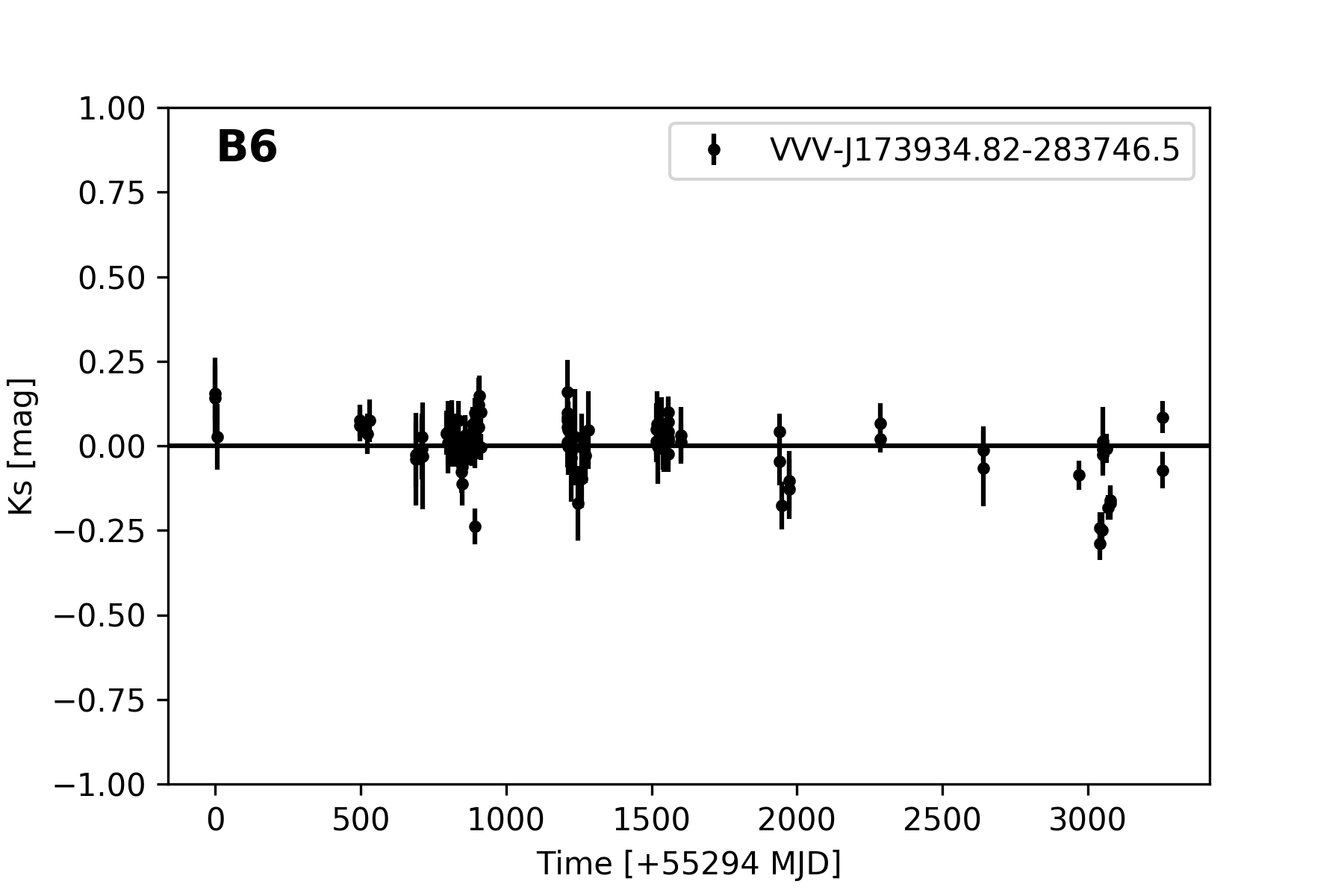}
\includegraphics[width=0.30\textwidth,height=0.25\textwidth]{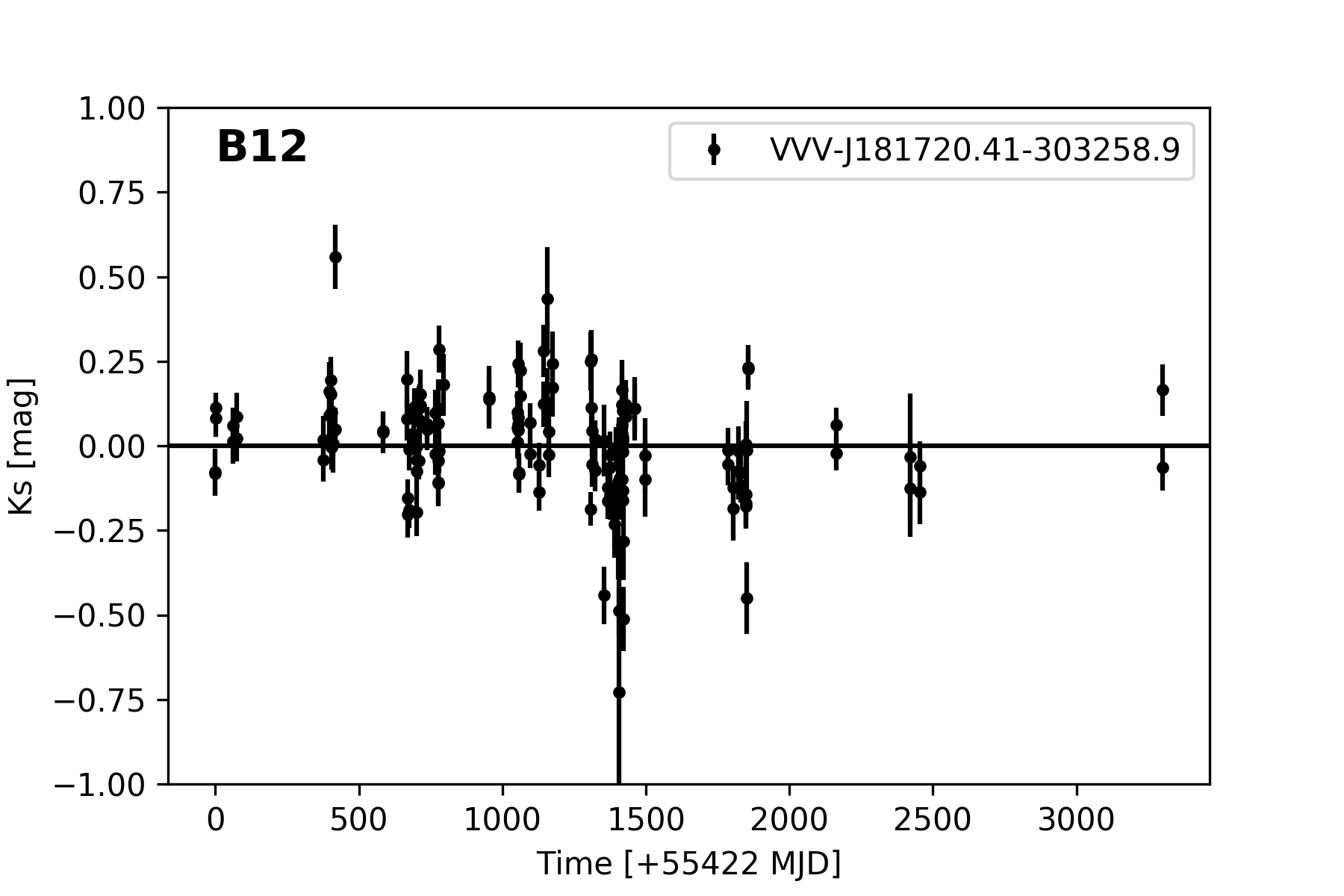}
\includegraphics[width=0.30\textwidth,height=0.25\textwidth]{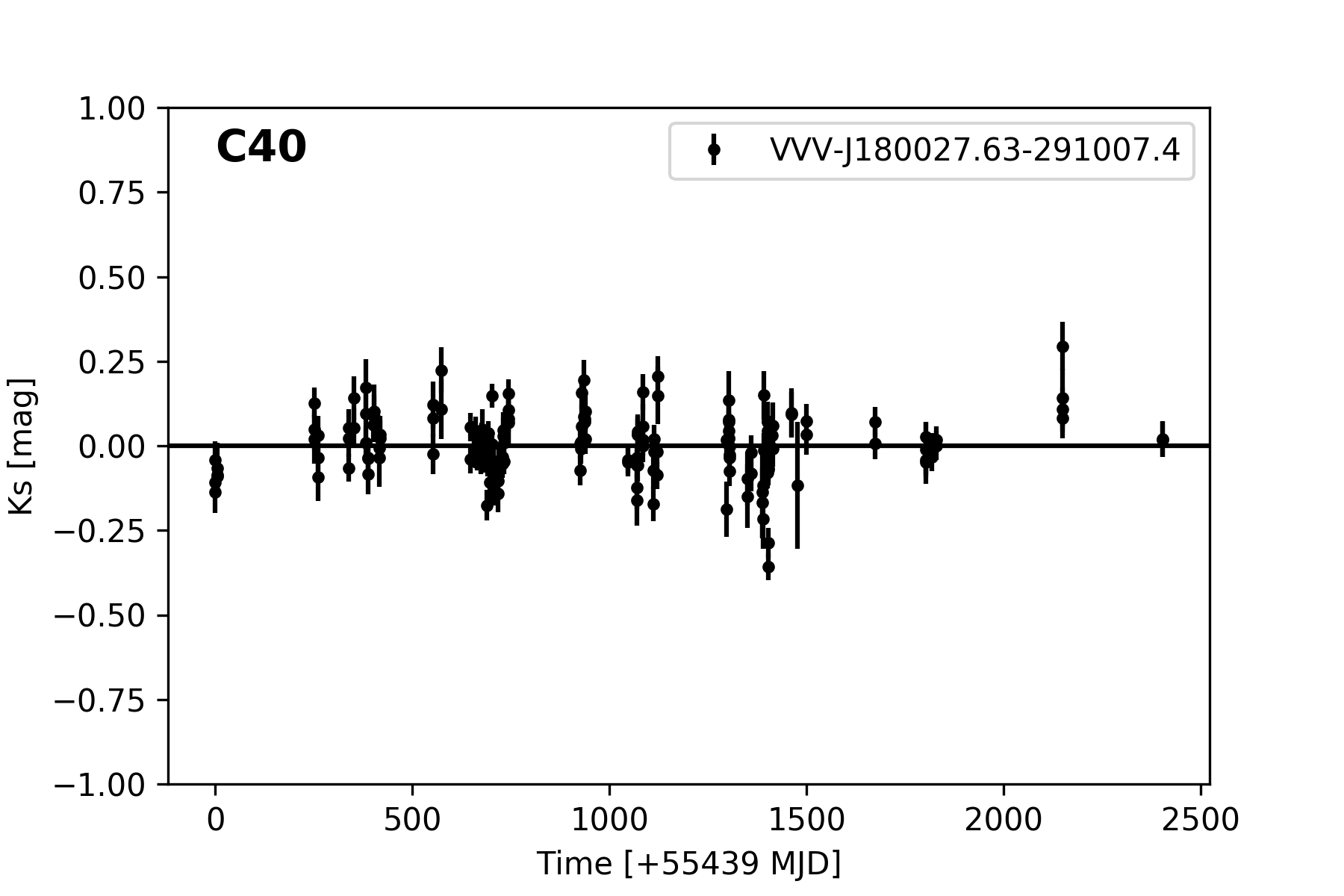}
\includegraphics[width=0.30\textwidth,height=0.25\textwidth]{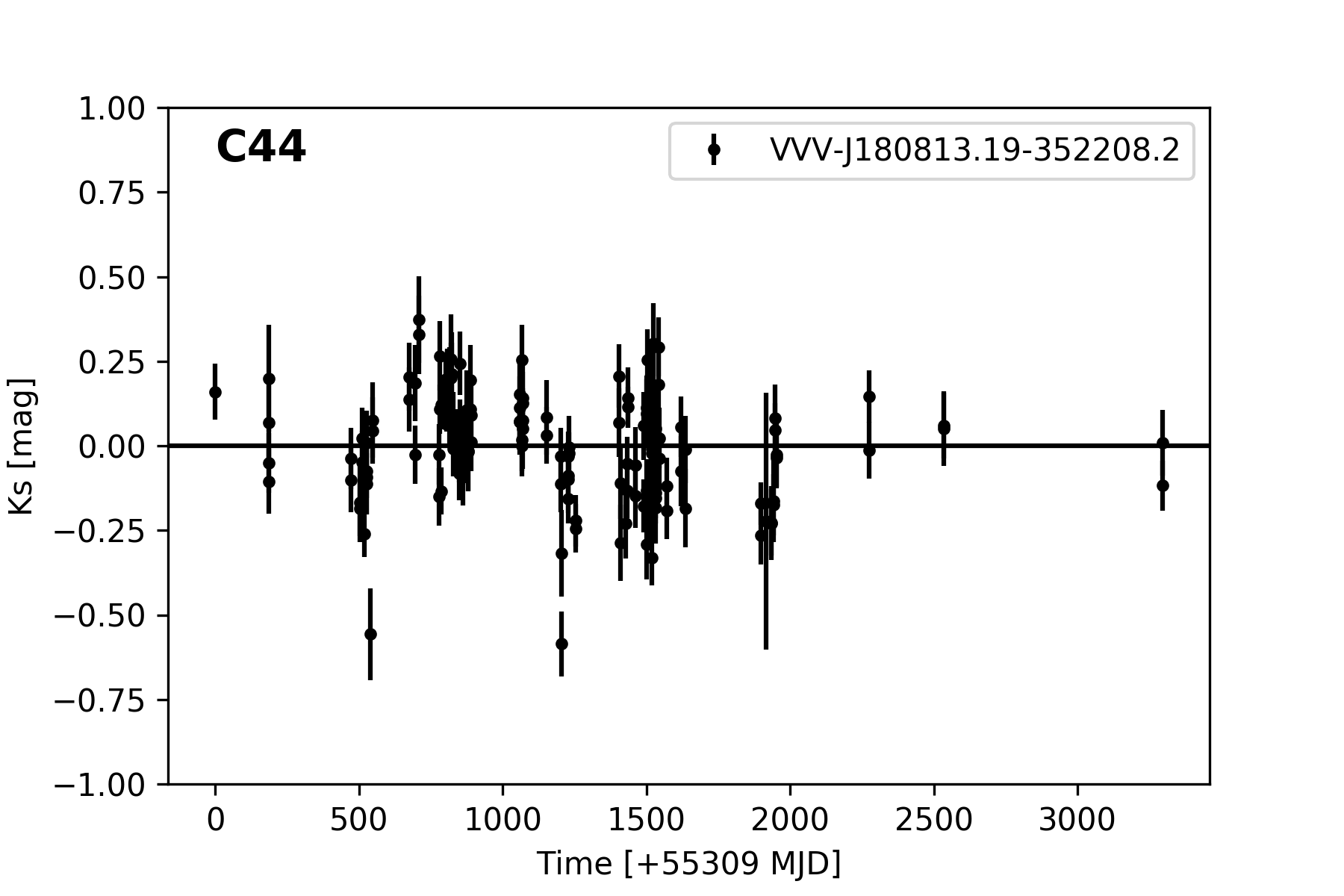}
\includegraphics[width=0.30\textwidth,height=0.25\textwidth]{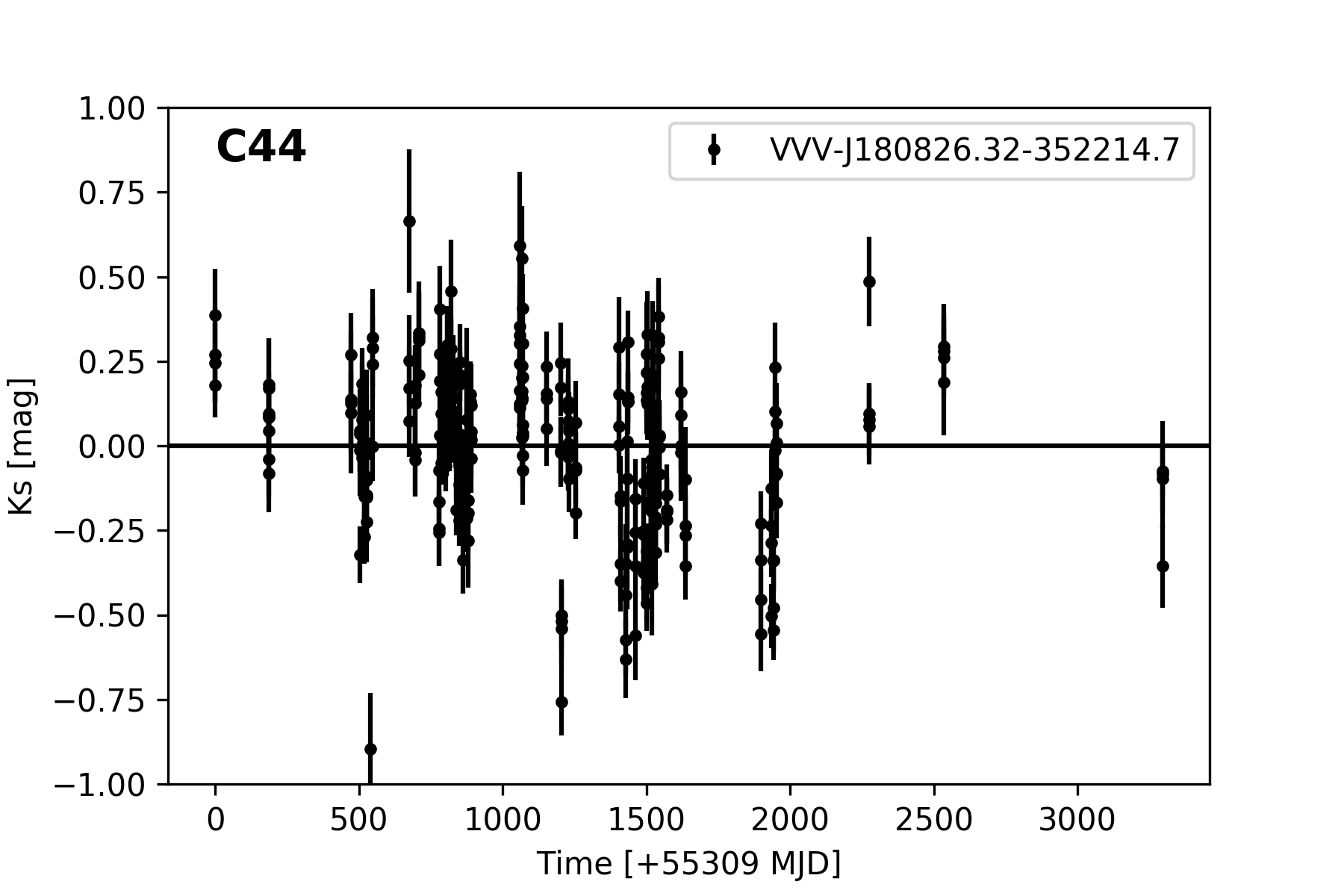}
\includegraphics[width=0.30\textwidth,height=0.25\textwidth]{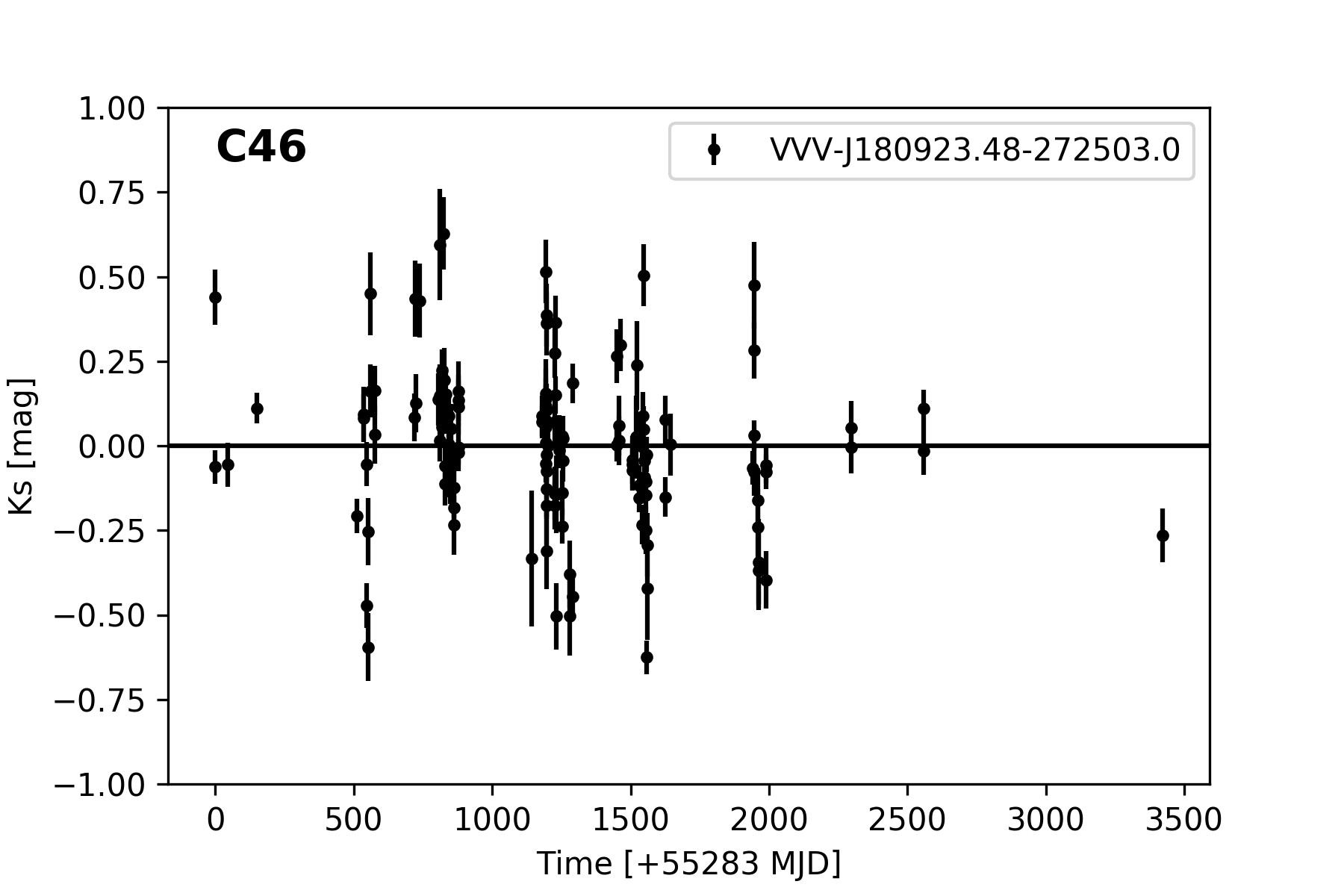}
\includegraphics[width=0.30\textwidth,height=0.25\textwidth]{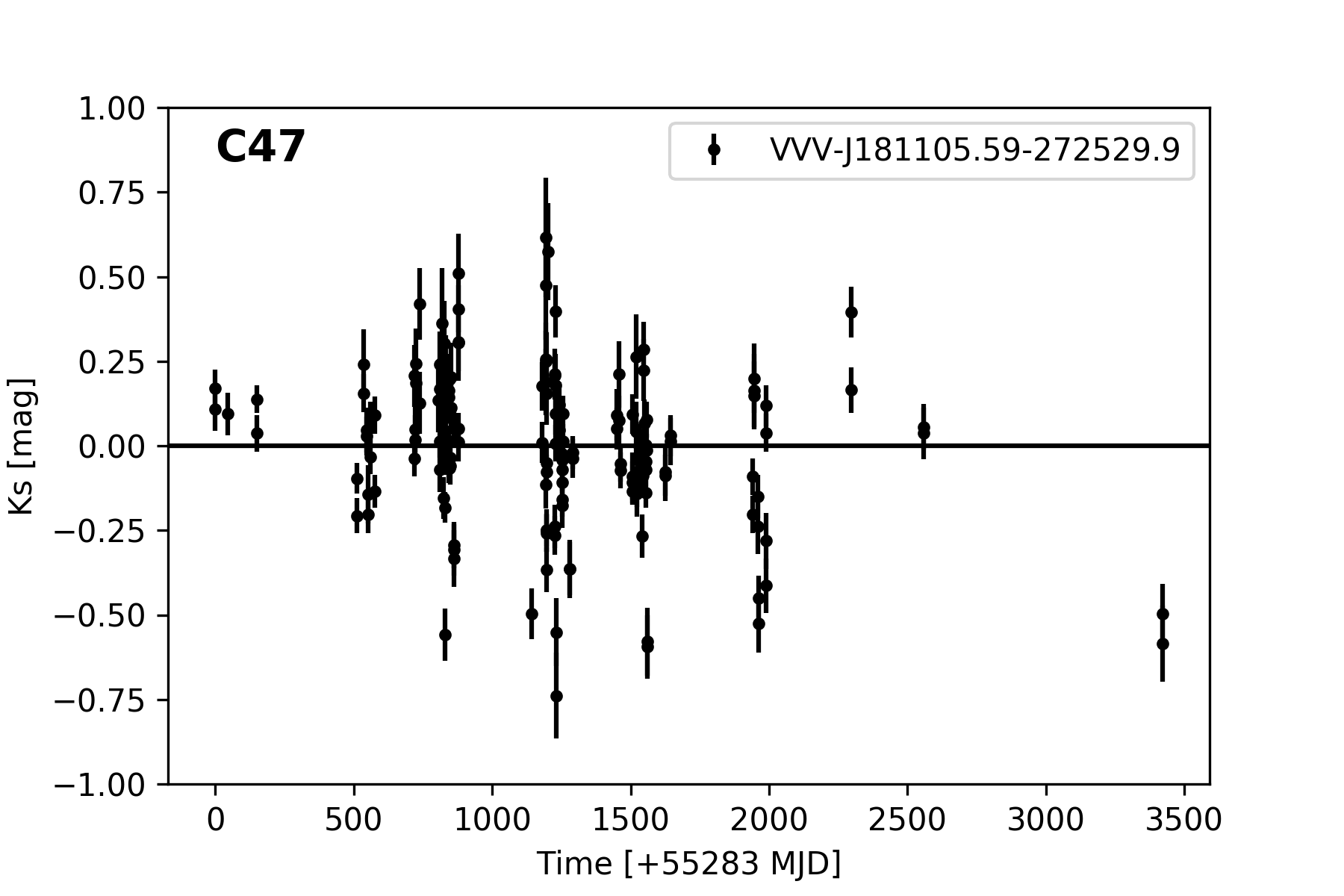}
\caption{K$_\mathrm{s}$ differential light curves of the VVV sources for the A, B and C subsamples.}
\label{fig:lightcurves}
\end{center}
\end{figure*}

\begin{figure*}
\begin{center}
 \contcaption{}
\includegraphics[width=0.30\textwidth,height=0.25\textwidth]{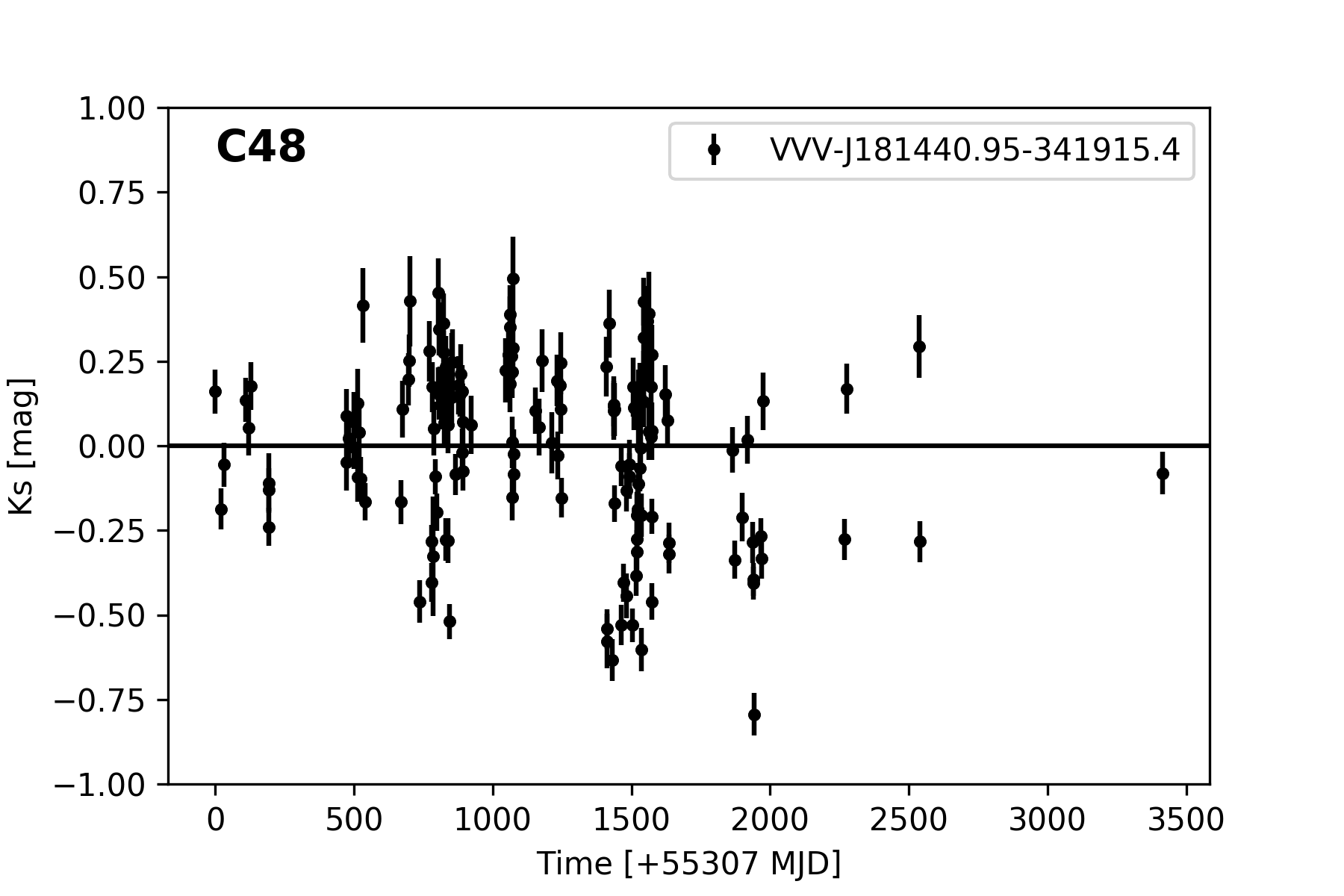}
\includegraphics[width=0.30\textwidth,height=0.25\textwidth]{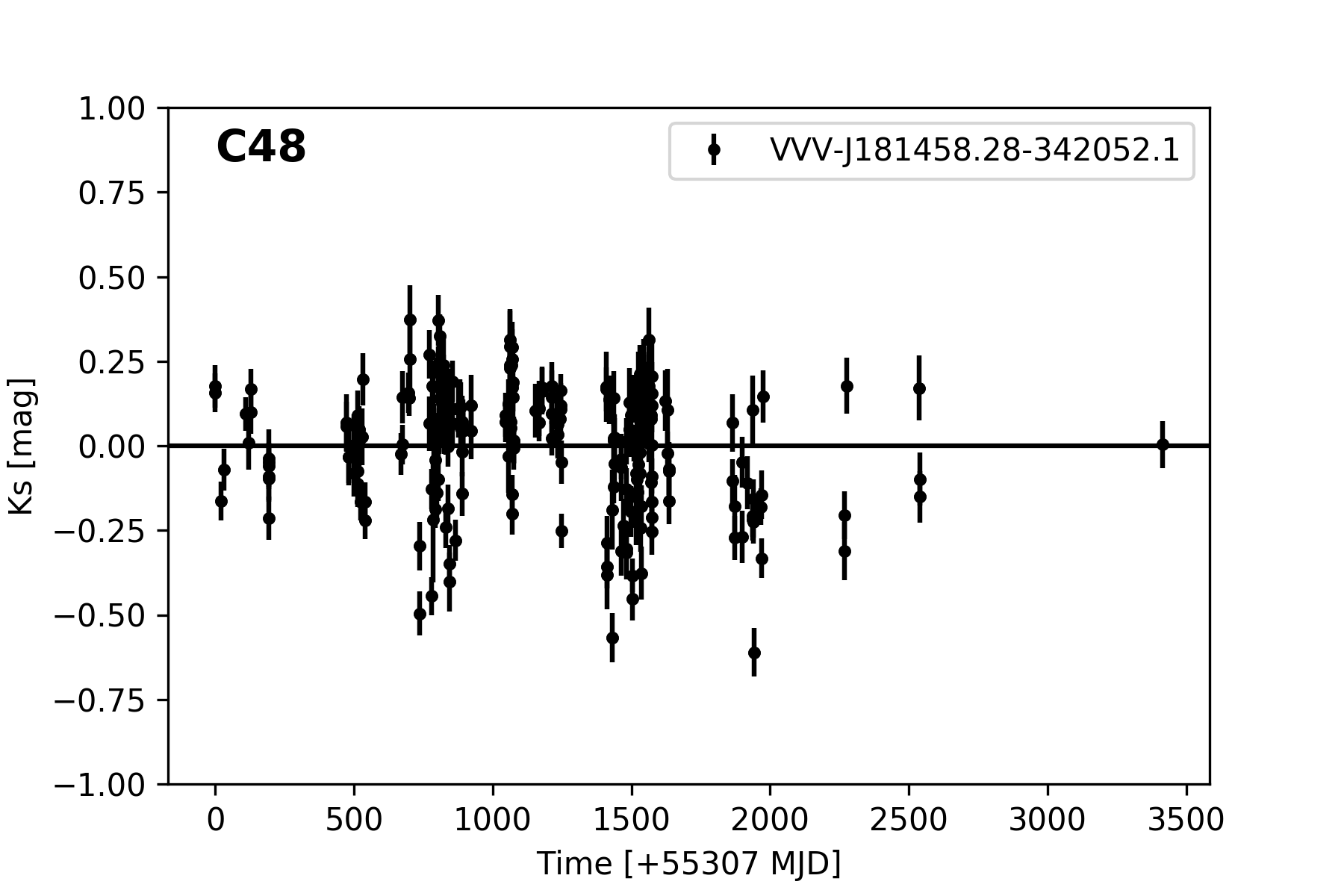}
\includegraphics[width=0.30\textwidth,height=0.25\textwidth]{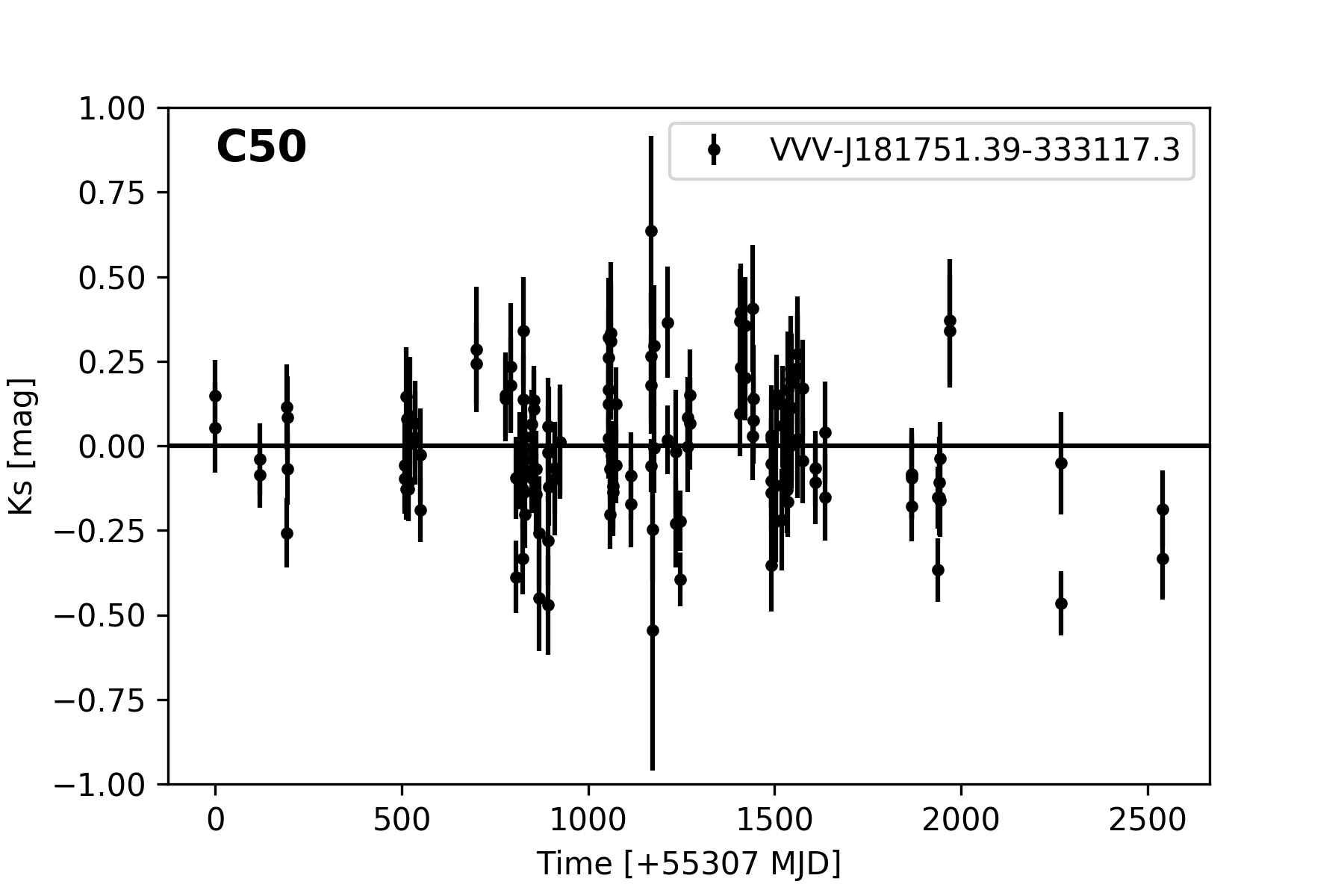}
\includegraphics[width=0.30\textwidth,height=0.25\textwidth]{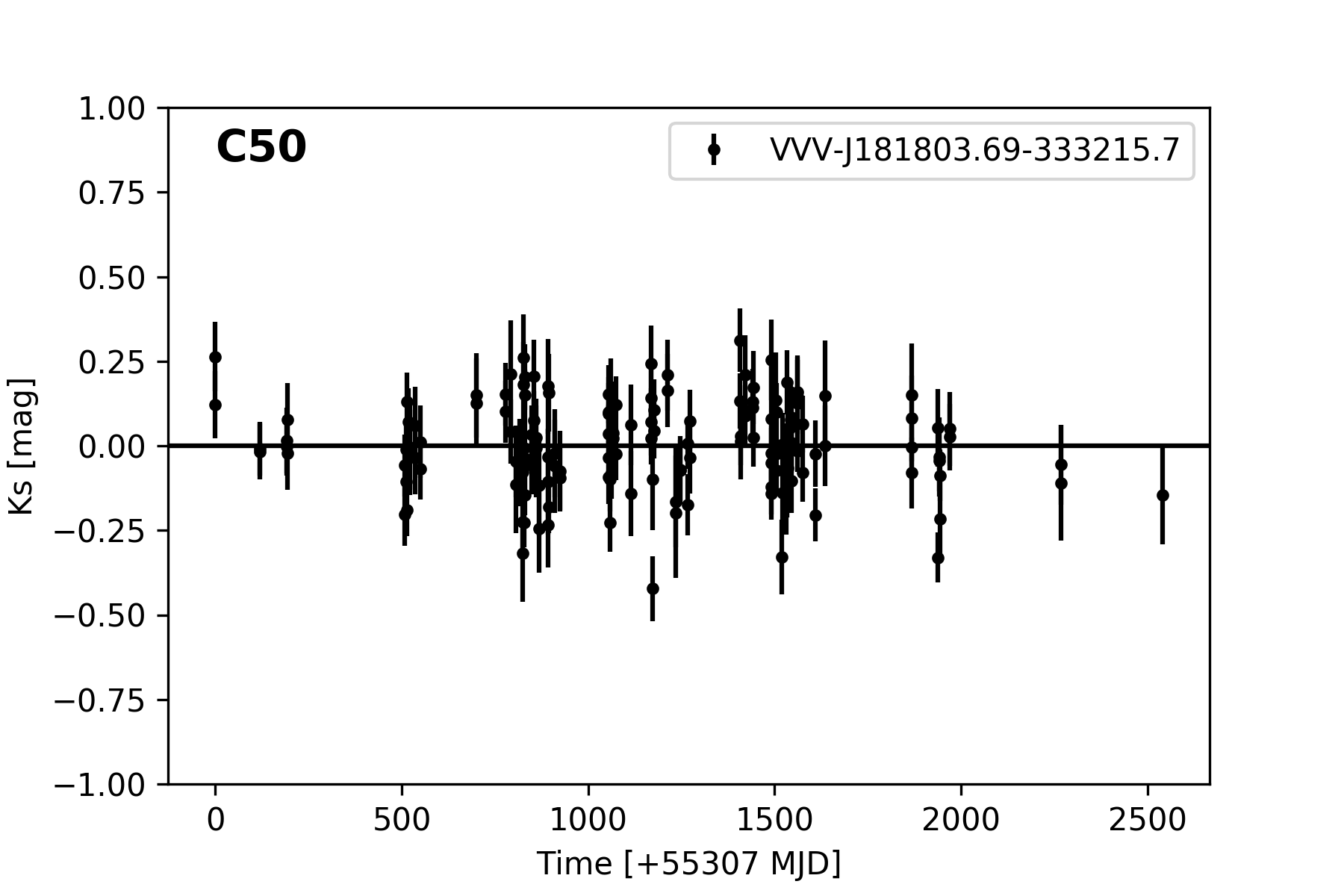}
\includegraphics[width=0.30\textwidth,height=0.25\textwidth]{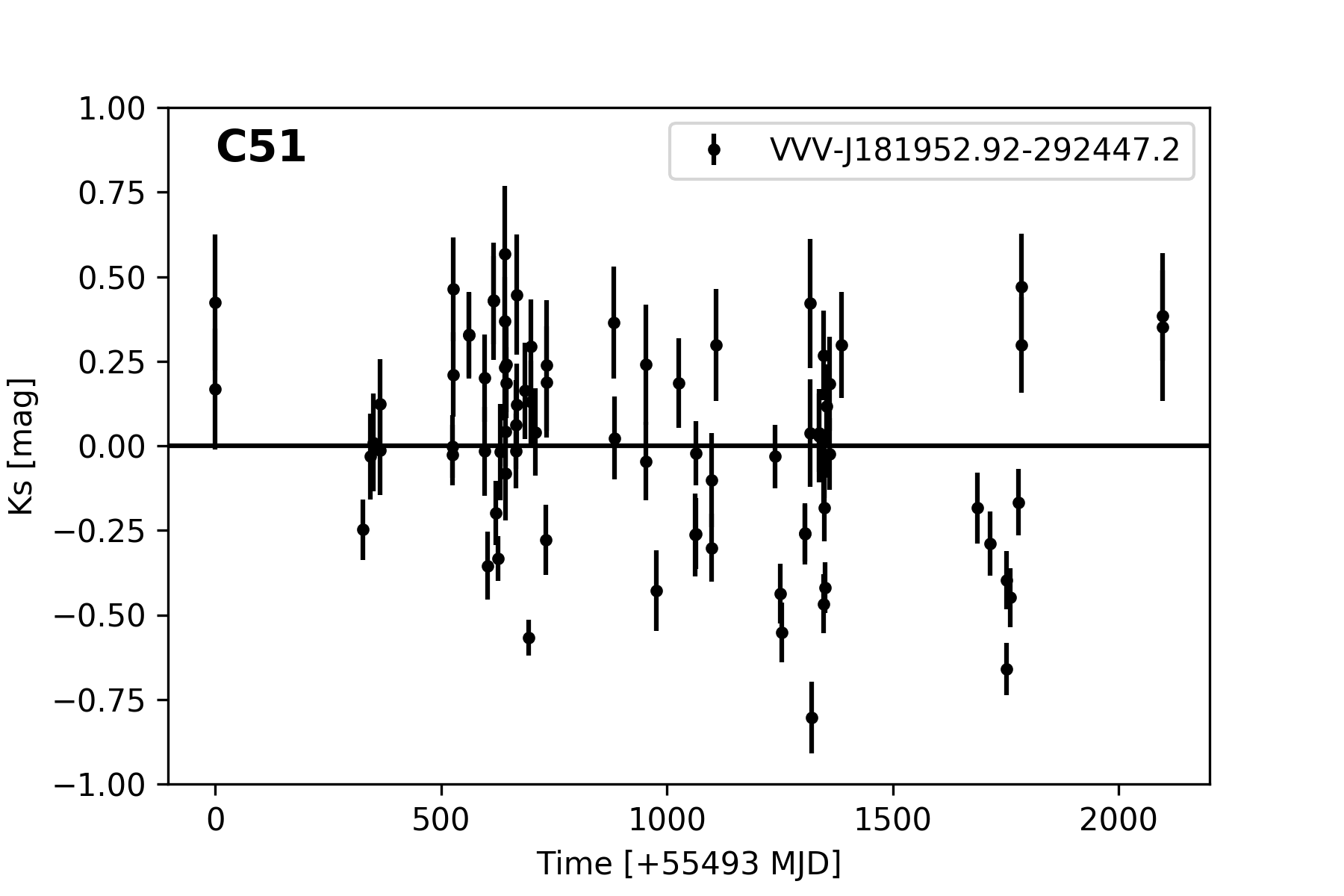}
\includegraphics[width=0.30\textwidth,height=0.25\textwidth]{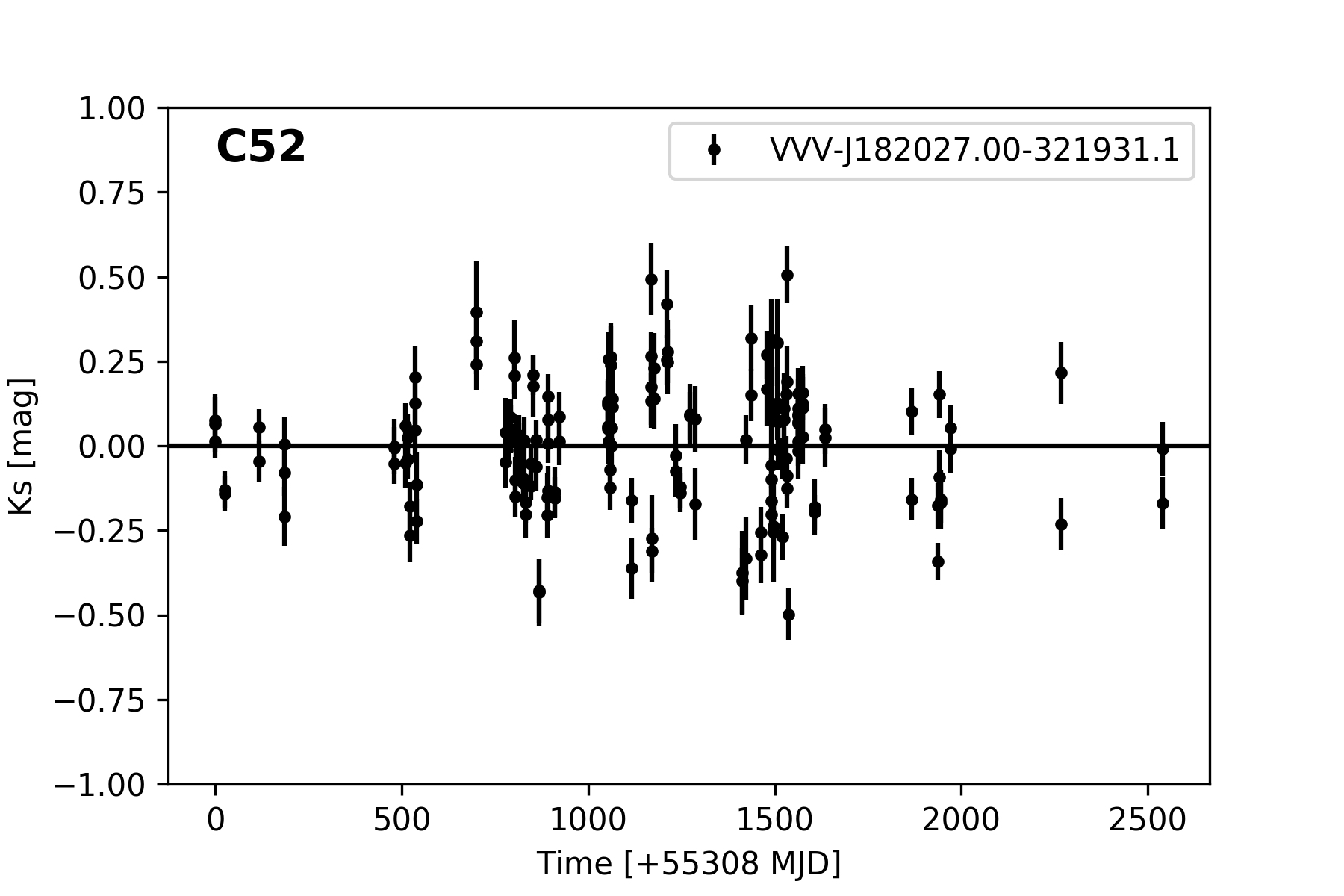}
\includegraphics[width=0.30\textwidth,height=0.25\textwidth]{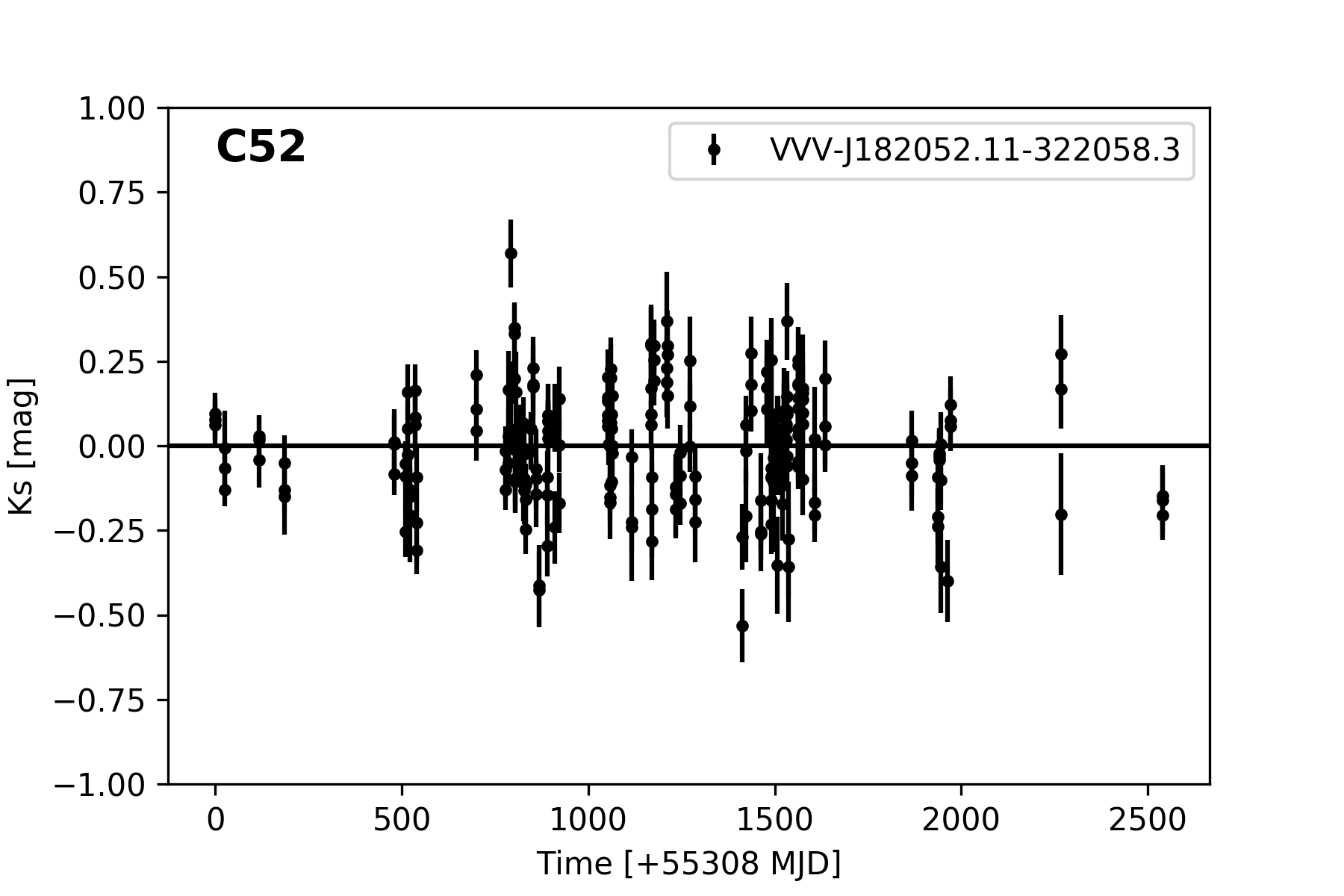}
\includegraphics[width=0.30\textwidth,height=0.25\textwidth]{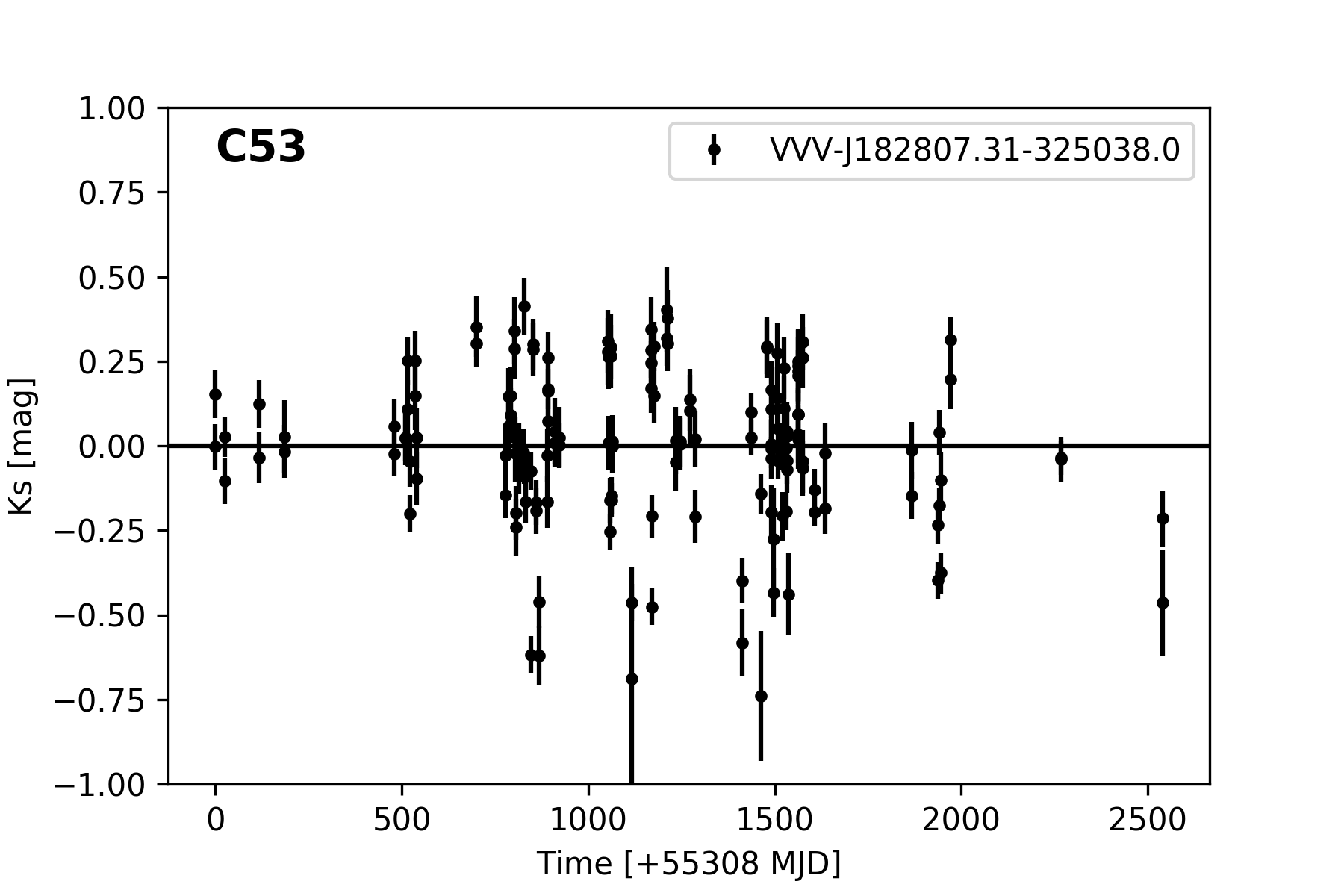}
\includegraphics[width=0.30\textwidth,height=0.25\textwidth]{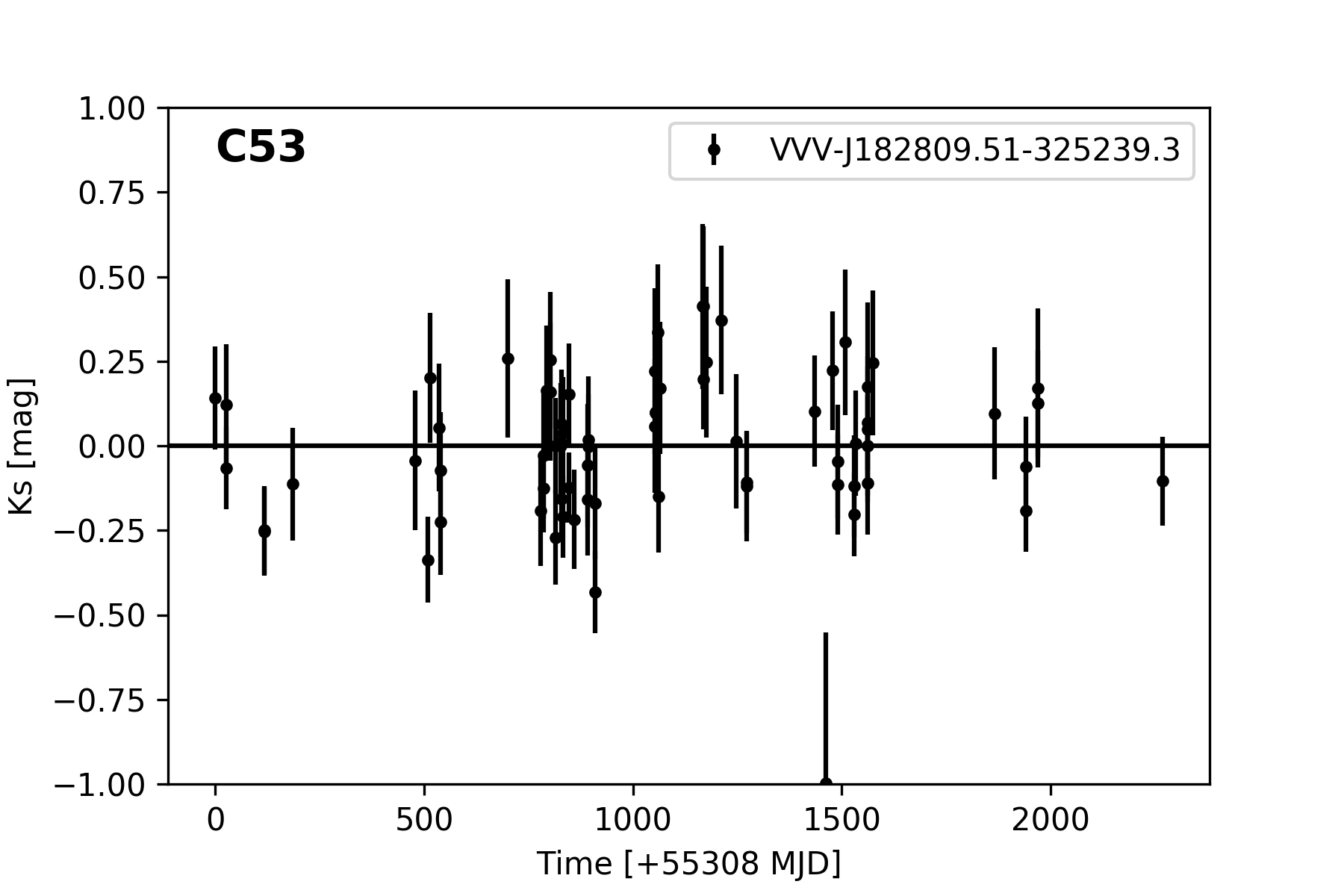}
\includegraphics[width=0.30\textwidth,height=0.25\textwidth]{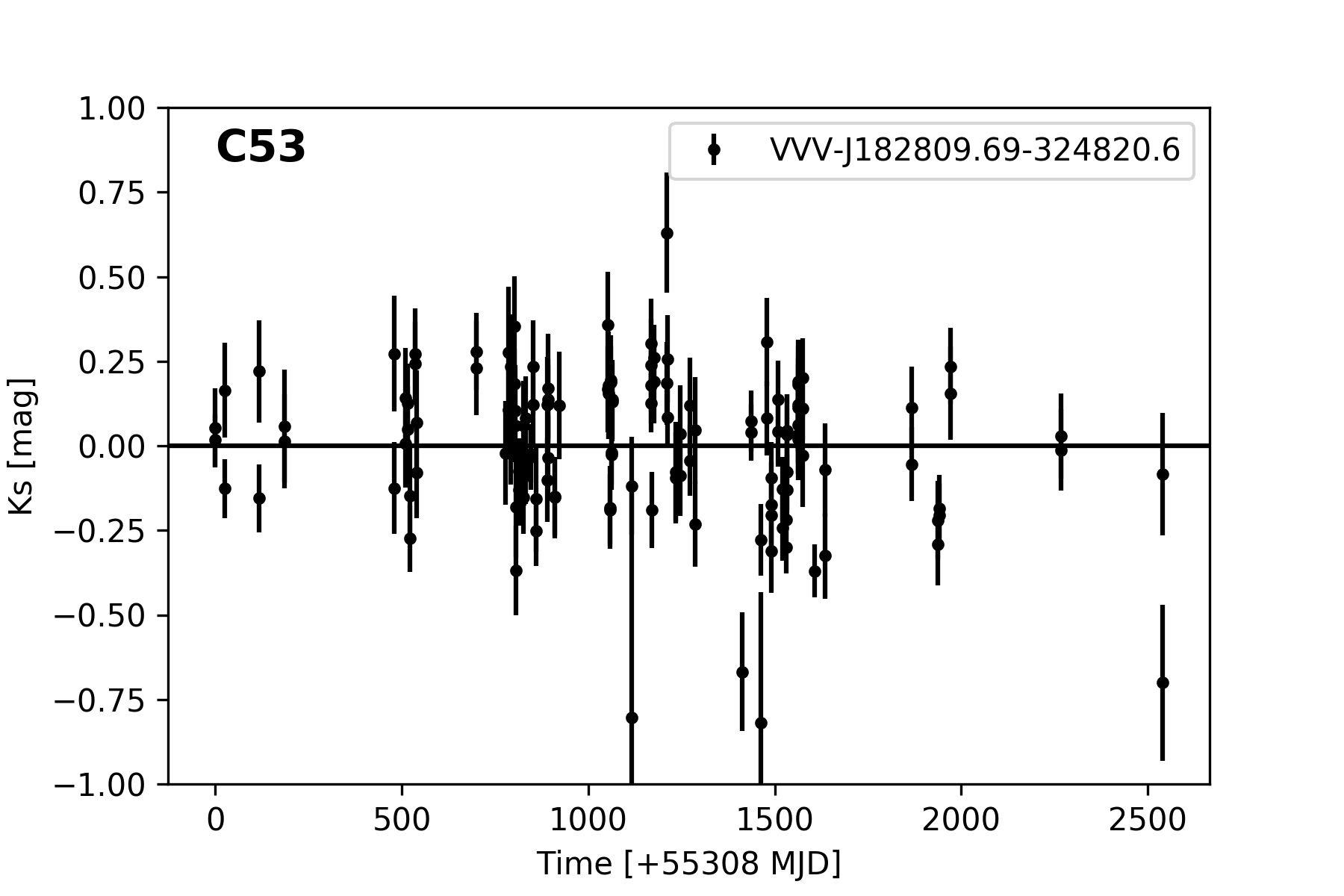}
\includegraphics[width=0.30\textwidth,height=0.25\textwidth]{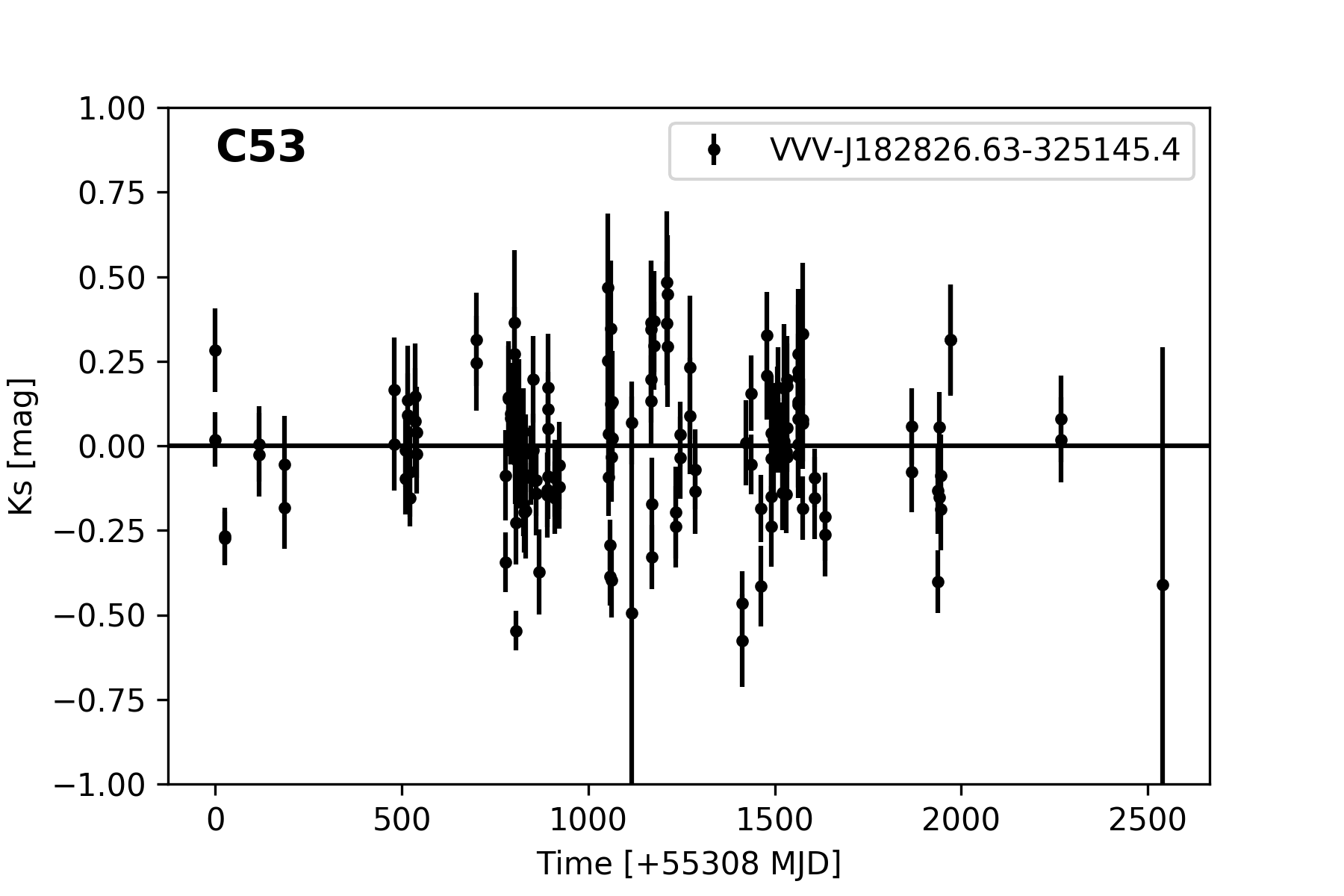}
\includegraphics[width=0.30\textwidth,height=0.25\textwidth]{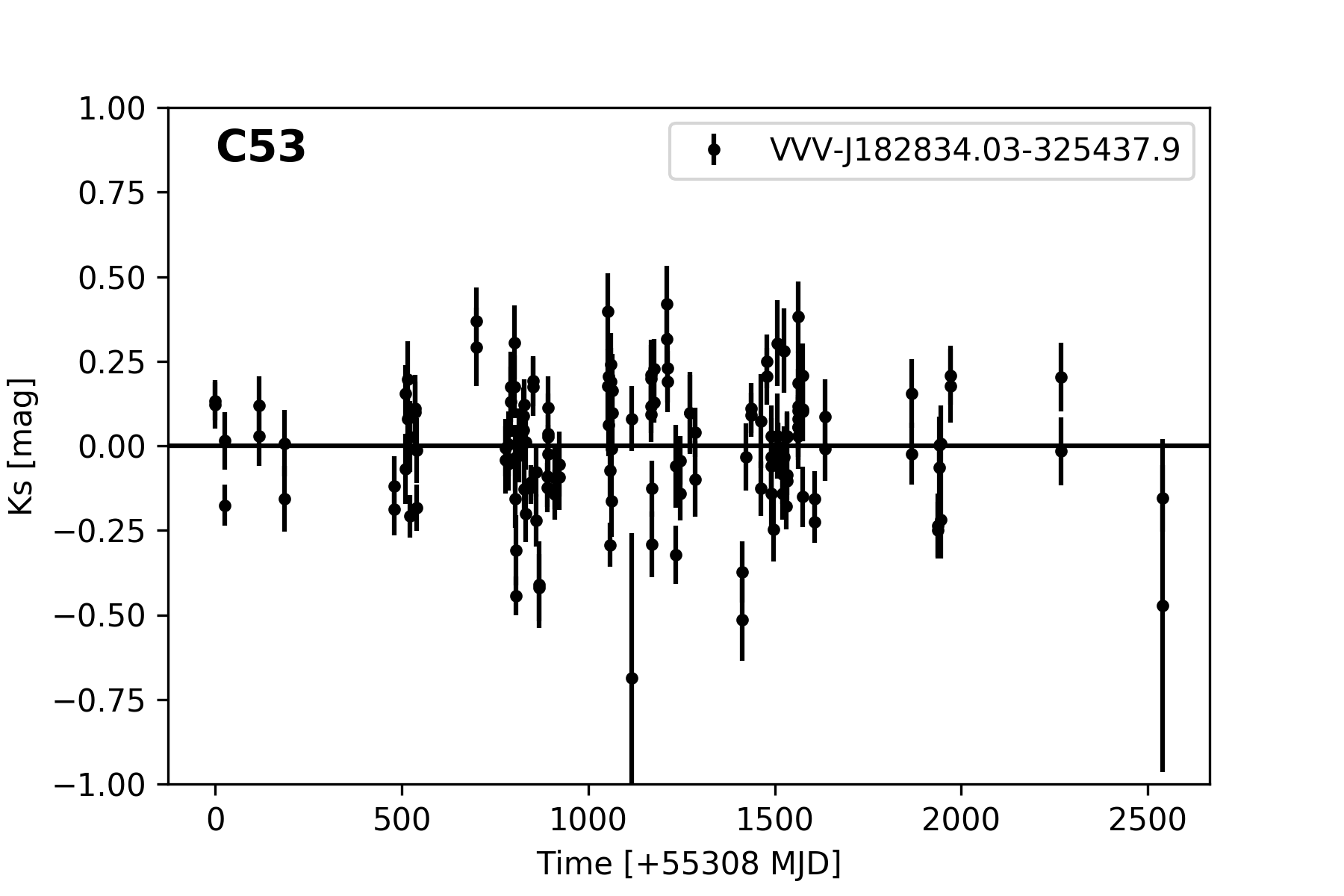}
\end{center}
\end{figure*}

\begin{table*}
\caption{K$_\mathrm{s}$ variability of the near-IR sources from the VVV survey. The internal identification of the 4FGL sources in the subsamples is listed in column (1); the VVV and the VIRAC2 identifications of the  sources is in columns (2) and (3), respectively; the mean K$_\mathrm{s}$ magnitude from VIRAC2 is in column (4); the fractional variability amplitude is in column (5); the range of days considered to derive the slope is in column (6); the absolute value of the slope of the variation is in column (7) and comments after the visual inspection of the sources is in column (8).}
\center
\begin{tabular}{|cccccccc|}
\hline 
 ID &  VVV ID     &  VIRAC2 ID   &  K$_\mathrm{s}$      & $\mathrm{\sigma_{rms}}$    & Range of days  &  Slope variation  & Visual classification \\ 
     &            &              &  [mag]   &    ($\times$ 100\%)       &  [mag/day]      & \\ 
\hline 
A9 & VVV-J175837.75-411002.2 & 14627836000675  & 17.02 $\pm$ 0.23 &	16.3    &	  600 - 1800  &  0.00013		&  Early-type galaxy \\ 
A9 & VVV-J175840.46-410757.8 & 14623740001044  & 17.70 $\pm$ 0.21 &	16.0    &	  1000 - 2700 &  0.00016		&  Early-type galaxy  \\
A9 & VVV-J175851.46-411016.0 & 14627837002111  & 15.89 $\pm$ 0.24 & 22.5    &	  0 - 1400    &  0.00016		&  Early-type galaxy\\
A9 & VVV-J175854.48-410927.7 & 14627837000010  & 16.20 $\pm$ 0.18 &	16.2    &	  300 - 1400  &  0.00013		&  Early-type galaxy \\
A12 & VVV-J181252.80-314443.5 & 13796388005480 & 16.24 $\pm$ 0.23 &	19.2    &	  0 - 1300    &   0.00015		&  Early-type galaxy \\
A12 & VVV-J181258.71-314346.7 & 13792293006414 & 15.30 $\pm$ 0.26 &	31.6    &	  0 - 1300    &   0.00020		&  Early-type galaxy \\
A12 & VVV-J181300.69-314505.6 & 13800485003177 & 16.22 $\pm$ 0.11 &	3.3     &	  300 - 1600  &   0.00008       &  faint galaxy \\ 
A13 & VVV-J183051.88-313056.9 & 13775960000711 & 16.36 $\pm$ 0.20 &	17.2    &	  1000 - 2600 &   0.00017		&  Early-type galaxy \\
\hline
B6  & VVV-J173934.82-283746.5 & 13501381012265  & 13.91 $\pm$ 0.08 &   5.0    &   0 - 3200 &   0.00005       &  point-like morphology  \\
B12 & VVV-J181720.41-303258.9 & 13685809002107  & 15.86 $\pm$ 0.16 &	14.0  &     250 - 1700  &  	0.00015     &  Early-type galaxy	 \\
\hline
C40  & VVV-J180027.63-291007.4 & 13554689002540 & 14.91 $\pm$ 0.10 & 6.8 & 250 - 2000 & 0.00004  & faint  galaxy \\
C44  & VVV-J180813.19-352208.2 & 14128151000796 & 16.28 $\pm$ 0.16 & 12.5 & 700 - 3300 & 0.00010 & Early-type galaxy  \\
C44  & VVV-J180826.32-352214.7 & 14128152000807 & 16.49 $\pm$ 0.23 & 20.8 & 0 - 3300 & 0.00010 & Early-type galaxy \\
C46  & VVV-J180923.48-272503.0 & 13382683008897 & 15.58 $\pm$ 0.23 & 20.3 & 700 - 3420 & 0.00013 & faint  galaxy  \\
C47  & VVV-J181105.59-272529.9 & 13386783001943 & 15.30 $\pm$ 0.23 & 22.1 & 0 - 3420 & 0.00012   & Early-type galaxy	\\
C48  & VVV-J181440.95-341915.4 & 14033961000873 & 15.81 $\pm$ 0.26 & 24.8 & 700 - 3420 & 0.00017 & Early-type galaxy	\\
C48  & VVV-J181458.28-342052.1 & 14033962001744 & 16.03 $\pm$ 0.18 & 16.1 & 700 - 2550 & 0.00012 & Early-type galaxy	\\
C50  & VVV-J181751.39-333117.3 & 13960242002550 & 16.55 $\pm$ 0.19 & 13.6 & 1000 - 2550 & 0.00016 & Late-type galaxy \\
C50  & VVV-J181803.69-333215.7 & 13960243002545 & 16.56 $\pm$ 0.13 & 8.0 & 0 - 2550 & 0.00002 & Early-type galaxy\\
C51  & VVV-J181952.92-292447.2 & 13575224000906 & 16.20 $\pm$ 0.32 & 32.1 & 0 - 2170 & 0.00017 & Early-type galaxy\\
C52  & VVV-J182027.00-321931.1 & 13849658000489 & 15.63 $\pm$ 0.20 & 15.6 & 1000 - 2550 & 0.00013 & Early-type galaxy\\
C52  & VVV-J182052.11-322058.3 & 13853755001577 & 15.99 $\pm$ 0.17 & 13.0 & 1000 - 2550 & 0.00011 & Early-type galaxy \\
C53  & VVV-J182807.31-325038.0 & 13898832000212 & 16.00 $\pm$ 0.23 & 21.7 & 1000 - 2550 & 0.00016 & Early-type galaxy \\
C53  & VVV-J182809.51-325239.3 & 13902928000442 & 17.04 $\pm$ 0.22 & 16.5 & 1000 - 2550 & 0.00020 & Late-type galaxy \\
C53  & VVV-J182809.69-324820.6 & 13894736002365 & 16.81 $\pm$ 0.22 & 17.4 & 1000 - 2550 & 0.00019 & Early-type galaxy \\
C53  & VVV-J182826.63-325145.4 & 13898833000294 & 16.54 $\pm$ 0.21 & 14.9 & 1000 - 2550 & 0.00014 & Late-type galaxy\\
C53  & VVV-J182834.03-325437.9 & 13907025000679 & 16.28 $\pm$ 0.19 & 15.0 & 1000 - 2550 & 0.00011 & Early-type galaxy\\
\hline 
\end{tabular}
\label{tab:variability}
\end{table*}

All the Fermi-LAT sources in the A subsample with AGN candidates are found in regions
with smaller interstellar extinctions (A$_\mathrm{Ks} < $ 0.15 mag). In the B subsample, there are only two Fermi-LAT sources associated with VVV AGN candidates: B12 has interstellar extinction lower than 0.10 mag and  B6 is in a region with high interstellar extinction. Hence, an interesting feature of B6 source's  colour–magnitude diagram is that the  K$_\mathrm{s}$ magnitudes are brightest compared to the other diagrams (Figure~\ref{f06.vvv}). In the C subsample most of the cases have A$_\mathrm{Ks} < $ 0.10 mag with the exception of C40, C46 and C47 with values from 0.17 to 0.28 mag approximately. For the other Fermi-LAT sources lying at higher interstellar extinction regions, we did not find any NIR nor MIR candidates.

Considering that some UGS have multiple candidates, it is crucial to establish criteria for prioritising the selection of the objects for follow-up observations. This selection process is based on additional criteria that includes magnitude, distance to the Fermi source, variability, interstellar extinction and  visual inspection. As a result, the priority candidate for the Fermi-LAT source A9 is VVV-J175851.46-411016.0, which is the brightest, closest, the most variable source and lowest interstellar extinction. For the Fermi-LAT source A12, the priority candidate is VVV-J181258.71-314346.7, using the same criteria mentioned above. Within the C subsample, the priority candidates are VVV-J180826.32-352214.7 for C44, VVV-J181440.95-341915.4 for C48, VVV-J181751.39-333117.3 for C50, VVV-J182052.11-322058.3 for C52 and VVV-J182807.31-325038.0 for C53.


\section{Summary}
\label{sec:final}
In this work we present criteria for selecting AGN candidates as counterparts to Fermi-LAT sources, based on NIR and MIR photometry
from the VVV and WISE surveys. We analysed a sample of 78 high energy $\gamma$-ray sources located at 
low Galactic latitudes without any previous source associations at any wavelength and lying in the footprint of the VVV survey. To start with, we divided the sample in three subsamples, considering the interstellar extinctions and
semi-major axis of the Fermi-LAT uncertainties.

We analysed photometric data from the VVV and WISE surveys, following the methodology reported by \cite{Pichel2020} to search for blazars and \cite{Baravalle2023} to identify AGN candidates.
The following colour cuts were
used to identify VVV AGN candidates associated to the UGS sample in the near-IR data: 
0.5 < (J-K$_\mathrm{s}$) < 2.5 mag; 0.5 < (H-K$_\mathrm{s}$) < 2.0 mag; 0.4 < (J-H) < 2.0 mag and 0.2 < (Y-J) < 2.0 mag. These sources are located in specific regions in the NIR CCD, clearly separated from stars and other extragalactic sources. 
Upon visual inspection, we removed the contaminated sources such as those with nearby bright stars or stellar associations.

We then selected 27 VVV AGN candidates within 14 Fermi-LAT positional uncertainties ellipses using the VVV survey.  These objects satisfy the colour cuts and also visually look as a galaxy or have point-like morphology. We have also explored the light curves of all sources reported in Table~\ref{tab:variability} and applied the fractional variability amplitude and the slope of variation in the  K$_\mathrm{s}$ passband. In general, most of the candidates show variability $\mathrm{\sigma_{rms}} > $  12 and slopes in agreement with the  limits defined by \citet{Cioni2011}. These results suggest the presence of type-1 AGN. However, there are four objects with low variability $\mathrm{\sigma_{rms}} < $ 8.0 and smaller slopes that might not be ruled out.  
We also found 2 blazar candidates in the regions of 2 Fermi-LAT sources using WISE data. There is no match between VVV and WISE candidates.

The combination of YJHK$_\mathrm{s}$ colours and K$_\mathrm{s}$ variability criteria have been useful for AGN selection, including its use in identifying counterparts to Fermi-LAT $\gamma$-ray sources.
Finally, we aim to perform NIR spectroscopic observations to confirm the extragalactic nature of the AGN candidates reported here. 
Particularly useful would be the data provided by Vera C. Rubin Observatory Legacy Survey of Space and Time \citep{LSST2009} and the eROSITA X-ray telescope \citep{Brunner2022} in order to complement this study.

\section*{Acknowledgements}
We want to thank the referee for useful comments and suggestions which has helped to improve this paper.
This work was partially supported by Consejo de Investigaciones Cient\'ificas y T\'ecnicas (CONICET), Secretar\'ia de Ciencia y T\'ecnica de la Universidad Nacional de C\'ordoba (SecyT) and Secretar\'ia de Ciencia y T\'ecnica de la Universidad Nacional de San Juan. 
D.M gratefully acknowledges support from the ANID BASAL projects ACE210002 and FB210003, from Fondecyt Project No. 1220724, and from CNPq/Brazil Project 350104/2022-0. The authors gratefully acknowledge data from the ESO Public Survey program IDs 179.B-2002 and 198.B-2004 taken with the VISTA telescope, and products from the Cambridge Astronomical Survey Unit (CASU).
We also thank Rom\'an Vena Valdarenas for helping us to improve the Figures. 
This research has made use of the VizieR catalogue access tool, CDS, Strasbourg, France (DOI: 10.26093/cds/vizier). The original description of the VizieR service was published in 2000, A\&AS 143, 23.

\section{Data availability}

The data underlying this article are available in \url{https://vvvsurvey.org/} and \url{https://www.nasa.gov/mission/wise}.



\bibliographystyle{mnras}
\bibliography{source} 

\bsp	
\label{lastpage}
\end{document}